%% file: main.tex
\documentclass[cis,noinfo,nocrop]{ipart_v1}

\firstpage{000}
\Vol{00}
\Issue{0}
\Year{2020}

\usepackage{graphicx}
\graphicspath{{figures/}}
\usepackage{amsmath,amssymb,dsfont,verbatim}
\usepackage{array}
\usepackage{caption}
\usepackage{subcaption}
\usepackage{import}
\usepackage{url}
\usepackage{xcolor}
\usepackage{booktabs} 
\usepackage{tikz}
\usetikzlibrary{mindmap}

\newif\ifarx

\subimport{custom/}{operators.tex}

\subimport{custom/}{environments.tex}

\begin{document}

\title[Second-Order Guarantees in Nonconvex Optimization]{Second-Order Guarantees in Centralized, Federated and Decentralized Nonconvex Optimization}

\author[S.~Vlaski and A.~H.~Sayed]{Stefan Vlaski and Ali H. Sayed \\ \vspace{10pt} In honor of Professor Thomas Kailath on the occasion of his 85th birthday\blfootnote{The authors are with the Institute of Electrical Engineering, \'Ecole Polytechnique F\'ed\'erale de Lausanne. Emails: \{stefan.vlaski, ali.sayed\}@epfl.ch.}
}

\begin{abstract}
    Rapid advances in data collection and processing capabilities have allowed for the use of increasingly complex models that give rise to nonconvex optimization problems. These formulations, however, can be arbitrarily difficult to solve in general, in the sense that even simply verifying that a given point is a local minimum can be NP-hard~\cite{Murty87}. Still, some relatively simple algorithms have been shown to lead to surprisingly good empirical results in many contexts of interest. Perhaps the most prominent example is the success of the backpropagation algorithm for training neural networks. Several recent works have pursued rigorous analytical justification for this phenomenon by studying the structure of the nonconvex optimization problems and establishing that simple algorithms, such as gradient descent and its variations, perform well in converging towards local minima and avoiding saddle-points. A key insight in these analyses is that gradient perturbations play a critical role in allowing local descent algorithms to efficiently distinguish desirable from undesirable stationary points and escape from the latter. In this article, we cover recent results on second-order guarantees for stochastic first-order optimization algorithms in centralized, federated, and decentralized architectures.
\end{abstract}

\maketitle

\section{Learning through Optimization}\label{sec:learning_through_optimization}
A key desirable feature of automated learning algorithms is the ability to learn models directly from data with minimal need for direct intervention by the designer. This is generally achieved by parameterizing a family of models of sufficient explanatory power through a set of parameters \( w \in \mathds{R}^M \) and subsequently searching for the choice of \( w^o \) that fits the data ``well'', in the sense that:
\begin{equation}\label{eq:centralized_expected}
  w^o \triangleq \argmin_{w \in \mathds{R}^M} \mathds{E}_{\boldsymbol{x}} Q(w; \boldsymbol{x} )
\end{equation}
In this formulation, the \emph{loss function} \( Q(w; \boldsymbol{x}) \) denotes a \emph{measure of fit} of the model \( w \) for the random data \( \boldsymbol{x} \). Hence, the desired model \( w^o \) is defined as the one that results in the smallest \emph{expected} risk, where the expectation is taken with respect to the distribution of the data \( \boldsymbol{x} \). As we illustrate in a number of examples in the sequel, a vast majority of inference problems fit into the general framework~\eqref{eq:centralized_expected}.
\begin{example}[Loss functions for supervised learning]\label{ex:loss}
  In the supervised learning setting, the data \( \boldsymbol{x} \triangleq \left\{ \boldsymbol{h}, \boldsymbol{\gamma} \right\} \) can be decomposed into a feature vector \( \h \in \mathds{R}^{M_1} \) and a label \( \boldsymbol{\gamma} \). When the target variable \( \boldsymbol{\gamma} \in \mathds{R}^{M_2} \) is continuous, this is typically an estimation problem with the objective being to construct an \emph{estimator} \( \widehat{\gamma}(w; \boldsymbol{h}) \) such that the error \( \widehat{\gamma}(w; \boldsymbol{h}) - \boldsymbol{\gamma} \) is small in some sense with high probability. One popular choice for the loss function \( Q(w; \boldsymbol{h}, \boldsymbol{\gamma}) \) in this case is the squared error loss:
  \begin{equation}\label{eq:ls_loss}
    Q(w; \boldsymbol{h}, \boldsymbol{\gamma}) = {\left \| \widehat{\gamma}(w; \boldsymbol{h}) - \boldsymbol{\gamma} \right \|}^2
  \end{equation}
  On the other hand, when $\boldsymbol{\gamma}$ is scalar and discrete such as the binary case \( \boldsymbol{\gamma} \in \{ -1, 1 \} \), the problem becomes a (binary) classification problem, with the objective being to find a classifier \( \widehat{\gamma}(w; \boldsymbol{h}) \) such that with high probability \( \mathrm{sign} \left\{ \widehat{\gamma}(w; \boldsymbol{h}) \right\} = \boldsymbol{\gamma} \). An example of a popular choice for the loss function in this case is the logistic loss:
  \begin{equation}\label{eq:logistic_loss}
    Q(w; \boldsymbol{h}, \boldsymbol{\gamma}) = \log \left( 1 + e^{-\boldsymbol{\gamma} \widehat{\gamma}(w; \boldsymbol{h})} \right)
  \end{equation}\hfill\qed%
\end{example}
We note that while the choice of the loss function is generally informed by the distribution of the target variable \( \boldsymbol{\gamma} \), such as whether it is continuous or discrete, we still need to specify the dependence of \( Q(w; \boldsymbol{h}, \boldsymbol{\gamma}) \) on \( w \). Since in both examples~\eqref{eq:ls_loss} and~\eqref{eq:logistic_loss}, the loss depends on \( w \) through \( \widehat{\gamma}(w; \boldsymbol{h}) \), we can describe this dependence by parameterizing \( \widehat{\gamma}(w; \boldsymbol{h}) \) through \( w \).

\begin{example}[Modeling for supervised learning]\label{ex:modeling}
  The most immediate parametrization of \( \widehat{\gamma}(w; \boldsymbol{h}) \) corresponds to the set of linear mappings:
  \begin{equation}\label{eq:linear_model}
    \widehat{\gamma}(w; \boldsymbol{h}) \triangleq \boldsymbol{h}^{\mathsf{T}} w
  \end{equation}
  Combining the linear model~\eqref{eq:linear_model} with the quadratic loss~\eqref{eq:ls_loss} results in the minimum mean-square error estimator, while~\eqref{eq:linear_model} with~\eqref{eq:logistic_loss} leads to the logistic regression solution, both of which are convex optimization problems with efficient solution methods~\cite{Sayed14}. While convexity of the resulting problem~\eqref{eq:centralized_expected} is an appealing property to have, the evident drawback of the linear parametrization~\eqref{eq:linear_model} is its limited expressive power. Only mappings that correspond to linear combinations of the elements of the feature vector are captured by~\eqref{eq:linear_model}, while non-linear interactions are beyond the scope of this model. For this reason, recent years have seen an increased interest in the utilization of neural networks, which are nested models of the form~\cite{lecun15}:
  \begin{equation}\label{eq:neural_model}
    \widehat{\gamma}(w; \boldsymbol{h}) \triangleq W_L \sigma\left( W_{L-1} \sigma\left( \ldots \sigma\left( W_1 \boldsymbol{h} \right) \right)  \right)
  \end{equation}
  where the \( \left\{ W_{\ell} \in \mathds{R}^{M_{\ell, 1} \times M_{\ell, 2}} \right\} \) denote matrices of appropriate dimensions and \( \sigma(\cdot) \) denotes an element-wise activation function (usually nonlinear in form). We can collect in \( w \) all parameters \( W_{\ell} \), i.e., \( w \triangleq \mathrm{col}\left\{ \mathrm{vec} \left\{ W_{\ell} \right\} \right\} \) and again recover an instance of~\eqref{eq:centralized_expected} for both the quadratic~\eqref{eq:ls_loss} and logistic~\eqref{eq:logistic_loss} losses. Models of the form~\eqref{eq:neural_model}, particularly for a suitable size \( L \) and dimensions \( M_{\ell, 1}, M_{\ell, 2} \) of hidden layers, are able to model well non-linear classification functions \( \widehat{\gamma}(w; \boldsymbol{h}) \). However, note that any choice \( L \ge 2 \) will generally result in a nonconvex loss surface~\eqref{eq:centralized_expected}. This necessitates the development of performance guarantees of algorithms for algorithms solving (1) under nonconvex environments.\hfill\qed%
\end{example}

\begin{example}[Unsupervised learning]\label{ex:unsupervised}
  Not all learning problems present themselves as supervised problems where the objective is to learn a mapping from feature to label. One such example is in the design of recommender systems where users are implicitly clustered and receive recommendations based on the preferences of ``similar'' other users. A popular approach on this setting revolves around matrix factorization~\cite{Koren09}. One such implementation results in:
  \begin{equation}\label{eq:low_rank}
    Q(w; \boldsymbol{x}) \triangleq \| \boldsymbol{X} - W_1 W_2^{\mathsf{T}} \|^2 + \rho \left( \| W_1 \|^2 + \|W_2\|^2 \right)
  \end{equation}
  where \( \boldsymbol{x} \triangleq \mathrm{vec}\left\{ \boldsymbol{X} \right\} \), \( w \triangleq \mathrm{col}\left\{ \mathrm{vec}\left\{ W_1 \right\}, \mathrm{vec}\left\{ W_2 \right\} \right\} \) and \( \rho > 0 \) denotes the regularization weights. The matrices \( W_1, W_2 \) are generally chosen to be tall, so that \( W_1 W_2^{\mathsf{T}} \) has low rank, and~\eqref{eq:low_rank} pursues a low-rank approximation of \( \boldsymbol{X} \).\hfill\qed%
\end{example}

\section{Centralized Stochastic Optimization}\label{sec:centralized}
From examples~\ref{ex:loss}--\ref{ex:unsupervised} we conclude that a large number of learning problems, including linear as well as non-linear regression and classification problems, and unsupervised formulations, can be recovered by specializing the general stochastic optimization problem~\eqref{eq:centralized_expected}. The task of designing an effective learning method then boils down to two related decisions: (a) the choice of the learning architecture, which determines the form of the loss \( Q(w; \boldsymbol{x}) \), and (b) the choice of the optimization strategy, which given realizations of the random variable \( \boldsymbol{x} \) yields a high-quality estimate for \( w^o \). For the remainder of this article we will consider the architecture, and hence \( Q(w; \x) \), fixed and will focus on the latter challenge, namely providing performance guarantees for the quality of the estimate of \( w^o \) produced by the optimization algorithm for general nonconvex problems. We let:
\begin{equation}\label{eq:define_j}
  J(w) \triangleq \mathds{E}_{\boldsymbol{x}} Q(w; \boldsymbol{x} )
\end{equation}

\subsection{Notions of Optimality}
Loosely speaking, the objective of any (stochastic) optimization algorithm is to produce ``high-quality'' estimates for the minimizer \( w^o \) in~\eqref{eq:centralized_expected}. When the risk \( J(w) \triangleq \mathds{E}_{\boldsymbol{x}} Q(w; \boldsymbol{x} ) \) is strongly-convex there is little ambiguity in the quantification of the quality of an estimate, since for strongly-convex costs with constant \( \nu \) we have~\cite[Sec. 9.1.2]{Vandenberghe04}:
\begin{equation}\label{eq:strong_string}
  \frac{\nu}{2} \| w - w^o \|^2 \le J(w)-J(w^o) \le \frac{1}{2\nu} \|\nabla J(w)\|^2
\end{equation}
If the risk additionally has \( \delta \)-Lipschitz gradients, we similarly have~\cite[Sec. 9.1.2]{Vandenberghe04}:
\begin{equation}\label{eq:smooth_string}
  \frac{1}{2\delta} \|\nabla J(w)\|^2 \le J(w)-J(w^o) \le \frac{\delta}{2} \| w - w^o \|^2
\end{equation}
By inspecting these two inequalities we conclude that all three measures of optimality, namely the squared deviation from the minimizer \( \|w - w^o\|^2 \), the excess risk \( J(w) - J(w^o) \), and the squared gradient norm \( \|\nabla J(w)\|^2 \) are essentially equivalent up to constants that depend on the strong-convexity and Lipschitz parameters \( \delta \) and \( \nu \), respectively. This means that, as long as the problem is reasonably well-conditioned, meaning that the fraction \( \frac{\delta}{\nu} \) does not grow too large, the choice of the performance measure is not particularly relevant, since high performance in one measure necessarily implies high performance in both other measures. In other words, any point \( w \in \mathds{R}^M \) with a small gradient norm \( \| \nabla J(w) \|^2 \), for strongly-convex problems, will essentially be globally optimal in the sense that both the excess risk \( J(w) - J(w^o) \) and distance to the minimizer \( \|w-w^o\|^2 \) will be small.

In the nonconvex setting considered here, and hence in the absence of~\eqref{eq:strong_string}, this is no longer the case as we illustrate in the sequel.
\begin{definition}[\(O(\mu)\)-first-order stationarity]\label{def:first-order}
  A point \( w \in \mathds{R}^M \) is \(O(\mu)\)-first-order stationary if:
  \begin{equation}
    \|\nabla J(w)\|^2 \le O(\mu)
  \end{equation}
  These points are technically only \emph{approximately} first-order stationary, since exact first-order stationarity would require \( \nabla J(w) = 0 \). Since we generally refer to \( O(\mu) \)-first-order stationarity throughout this manuscript, we will drop ``approximate'' for convenience whenever it is clear from context.\hfill\qed%
\end{definition}
In light of relation~\eqref{eq:smooth_string}, for costs with \( \delta \)-Lipschitz gradients, \(O(\mu)\)-first-order stationarity is a \emph{necessary condition} to ensure \( J(w)-J(w^o) \le O(\mu) \) and \( \| w - w^o \|^2 \le O(\mu)\). However, unless the cost is assumed to additionally be strongly convex, Definition~\ref{def:first-order} \emph{is not sufficient} to guarantee that the point \( w \) has small excess risk \( J(w)-J(w^o) \) or small distance to the minimizer \( \| w - w^o \|^2 \), since establishing sufficiency requires~\eqref{eq:strong_string} which only holds for \emph{strongly-convex} costs. In fact, the set of \( O(\mu) \)-first-order stationary points for nonconvex risk functions includes the set of local minima, maxima as well as saddle-points.
Nevertheless, many studies of local descent algorithms in nonconvex environments establish performance guarantees by showing that the limiting points of the algorithm are approximately first-order stationary using variations of Definition~\ref{def:first-order} in both the single-agent and multi-agent settings~\cite{Nesterov98, Bertsekas00, Reddi16, DiLorenzo16, Tatarenko17,Lian17, Tang18, Wang18}. These results are reassuring, as first-order stationarity is a necessary condition for local optimality, and hence any algorithm that does not produce a first-order stationary point will necessarily not produce a point with small excess risk, or small distance to the minimizer. Nevertheless, these results cannot ensure that the limiting first-order stationary point does not correspond to a saddle-point, which have been identified as a bottleneck in many nonconvex problems of interest~\cite{Choromanska14}. This observation, following the works~\cite{Nesterov06, Ge15, Lee16} motivates us to consider {a} stronger notion of optimality.

{To formulate it, note that our objective is to converge towards points \( w \in \mathds{R}^M \) that are local minima and hence satisfy:
\begin{equation}\label{eq:minimum}
  J(w) \le J(w+\Delta w)
\end{equation}
for all small \( \Delta w \in \mathds{R}^M \). In other words, we would like to avoid approaching points \( w \) where there exists \( \Delta w \in \mathds{R}^M \) such that:
\begin{equation}
  J(w) > J(w+\Delta w)
\end{equation}
By introducing the second-order Taylor expansion around \( w \), we can write:
\begin{align}\label{eq:taylor}
  J(w) - J(w+\Delta w) \approx&\: -{\nabla J(w)}^{\T} \Delta w - {\Delta w}^{\T} \nabla^2 J(w) \Delta w \notag \\
  \approx&\: -{\Delta w}^{\T} \nabla^2 J(w) \Delta w
\end{align}
where we dropped the linear term \( {\nabla J(w)}^{\T} \Delta w \) since, at first-order stationary points, \( \nabla J(w) \approx 0 \). Hence, we shall say that \( w \) is \emph{second-order} locally optimal according to its \emph{second-order Taylor expansion} if, and only if,
\begin{equation}
  {\Delta w}^{\T} \nabla^2 J(w) \Delta w \ge 0
\end{equation}
This requirement is equivalent to \( \lambda_{\min} \left( \nabla^2 J(w) \right) \ge 0 \). We emphasize that \( w \) is \emph{second-order} locally optimal, since expression~\eqref{eq:taylor} is only an approximation of \( J(w) - J(w+\Delta w) \) based on derivatives up to second-order. Therefore, approaching points \(w\) where \(\lambda_{\min}(\nabla^2 J(w))\geq 0\) is desirable. Another way to see this is to note that we also have from~\eqref{eq:taylor}:
\begin{align}\label{eq:taylor_eigen}
  J(w) \approx&\: J(w+\Delta w) - {\Delta w}^{\T} \nabla^2 J(w) \Delta w \notag \\
  \le&\: J(w+\Delta w) - \lambda_{\min}\left( \nabla^2 J(w) \right) \|\Delta w\|^2
\end{align}
where equality holds whenever \( \Delta w \) is the eigenvector of \( \nabla^2 J(w) \) corresponding to \( \lambda_{\min}\left( \nabla^2 J(w) \right) \), i.e., \( \nabla^2 J(w) \Delta w = \lambda_{\min}\left( \nabla^2 J(w) \right) \Delta w \). It follows that whenever $\lambda_{\min}\left( \nabla^2 J(w)\right)$ is negative, the larger its magnitude is, the less locally optimal \( w \) is. In other words, points \( w \) with significantly negative \( \lambda_{\min}\left( \nabla J(w) \right) \) are highly undesirable limiting points of a local descent algorithm. Motivated by this discussion, we define the set of \(\tau\)-second-order stationary points. }
\begin{definition}[{\(\tau\)-}second-order stationarity]\label{def:second-order}
  A point \( w \in \mathds{R}^M \) is {\(\tau\)-}second-order stationary if it is \( O(\mu) \)-first-order stationary following Definition~\ref{def:first-order} and additionally, for some \( \tau > 0 \),
  \begin{equation}\label{eq:second-order}
    \lambda_{\min} \left\{ \nabla^2 J(w) \right\} > -\tau
  \end{equation}
  where \( \lambda_{\min} \left\{ \nabla^2 J(w) \right\} \) denotes the smallest eigenvalue of the Hessian matrix \( \nabla^2 J(w) \).\hfill\qed%
\end{definition}
We will be focusing on the case when \(\tau\) is small. Intuitively, points \(w\) that satisfy condition~\eqref{eq:second-order} are either local minima (e.g., when all eigenvalues of the Hessian matrix are positive) or they are weak saddle-points that are close to local minima (when the smallest eigenvalue is negative but only by a small amount). {Returning to~\eqref{eq:taylor_eigen}, we find that every \( \tau \)-second-order stationary point \( w \) satisfies:
\begin{equation}\label{eq:local_optimality}
  J(w) \le J(w+\Delta w) + \tau \|\Delta w\|^2
\end{equation}
Note that, as \( \tau \to 0 \), the definition of \( \tau \)-second-order stationarity corresponds to the definition of local optimality~\eqref{eq:minimum}. The freedom to set any \( \tau > 0 \), rather than requiring \( \tau \to 0 \), allows us to set an expectation of local optimality in the sense of~\eqref{eq:local_optimality}. This quantity does not appear as a parameter of any of the algorithms presented in this work, but does appear in the expressions on the convergence time (Theorems~\ref{TH:DESCENT_THROUGH_SADDLE_POINTS} and~\ref{TH:DESCENT_THROUGH_SADDLE_POINTS_DEC}) as \( O(1/{\tau}) \) meaning that a higher expectation of local optimality requires longer running time of the algorithms, which conforms with intuition. We conclude that, while for non-zero \( \tau \) not all \( \tau \)-second-order stationary points are locally optimal, any \( \tau \)-second-order stationary is \emph{almost} locally optimal for small \( \tau \) in the sense of~\eqref{eq:local_optimality}.}

{T}he set of second-order stationary points in Definition~\ref{def:second-order} is a subset of the set of first-order stationary points in Definition~\ref{def:first-order}. Every second-order stationary point is also first-order stationary, but the additional restriction~\eqref{eq:second-order} allows for the exclusion of certain, undesirable, stationary points {that do not satisfy~\eqref{eq:local_optimality}}, such as local maxima and saddle-points. {Specifically, by choosing \( \tau \) small enough, we are able to exclude any first-order stationary point where the smallest eigenvalue of the Hessian is negative and bounded away from zero. These points{, which correspond to the complement of Definition~\ref{def:second-order},} are {frequently} referred to as \emph{strict} saddle-points {in the literature} due to the requirement for the smallest eigenvalue to be \emph{strictly} negative.
\begin{definition}[{\(\tau\)}-strict saddle-points]\label{def:strict-saddle}
  A point \( w \in \mathds{R}^M \) is a {\(\tau\)-}strict saddle-point if it is \( O(\mu) \)-first-order stationary following Definition~\ref{def:first-order} and additionally:
  \begin{equation}\label{eq:strict-saddle}
    \lambda_{\min} \left\{ \nabla^2 J(w) \right\} \le -\tau
  \end{equation}
  Note that the only difference to Definition~\ref{def:second-order} is the reversal of inequality~\eqref{eq:second-order} to~\eqref{eq:strict-saddle}. As such, the set of \( \tau\)-strict saddle-points is precisely the complement of the set of \(\tau\)-second-order stationary points in the set of first-order stationary points.\hfill\qed%
\end{definition}
{Note that, depending on the choice of the parameter \( \tau \), not all saddle-points of the cost \( J(w) \) need to be \( \tau \)-strict saddle-points. If \( J(w) \) happens to have a saddle-point with \( - \tau \le \lambda_{\min}\left( \nabla^2 J(w) \right) < 0 \), then this particular saddle-point would not be \( \tau \)-strict , and in fact would fall under Definition~\ref{def:second-order} of a \( \tau \)-second-order stationary point. Nevertheless, so long as \( \tau \) is small, such saddle-points can intuitively be viewed as ``weak'' saddle-points, in the sense that they are \emph{almost} locally optimal according to~\eqref{eq:local_optimality}.}

{Under this formal definition, the set of strict saddle-points includes local maxima as well. In fact, if \emph{all} eigenvalues of \( \nabla^2 J(w) \) at a first-order stationary point \( w \) were bounded from above by \( -\tau \), then \( w \) would be a \emph{local maximum}. The set of strict saddle-points, however, is larger than the set of local maxima, since \emph{only one} eigenvalue of the Hessian at strict saddle-points is required to be bounded from above by \( -\tau \), while other eigenvalues are unrestricted. Hence, the incorporation of second-order information in the definition of stationarity allows us to distinguish between {\(\tau\)-}second-order stationary points and \(\tau\)-strict saddle-points and allows for the exclusion of points with significant local descent direction from the set of potentially optimal points. Furthermore, for many loss functions commonly found in machine learning, such as tensor decomposition~\cite{Ge15}, matrix completion~\cite{Ge16}, low-rank recovery~\cite{Ge17} and some deep learning formulations~\cite{Kawaguchi16}, \emph{all} saddle-points {and local maxima} have been shown to have a significant negative eigenvalue in the Hessian, and can hence be excluded from the set of second-order stationary points for sufficiently small{, but finite}, \( \tau \). {For such risk functions, \emph{all} \( \tau \)-second-order stationary points for some small, but finite, \( \tau \) correspond to local, or even global, minima.}
\begin{figure}
\centering
\begin{subfigure}[b]{.3\textwidth}
  \centering
  \includegraphics[width=\linewidth]{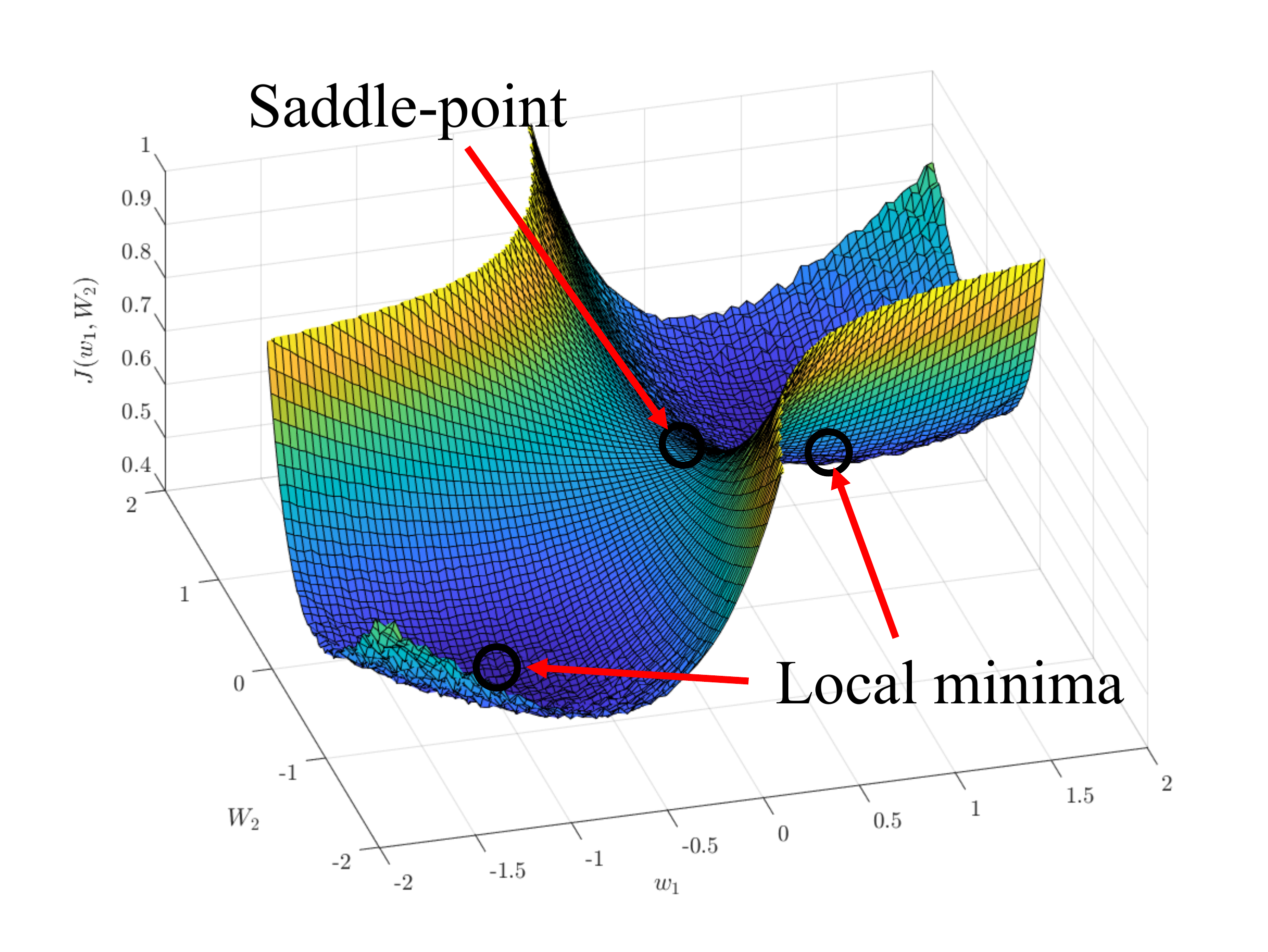}
  \caption{{Sample risk surface.}}
\end{subfigure}%
\begin{subfigure}[b]{.3\textwidth}
  \centering
  \includegraphics[width=\linewidth]{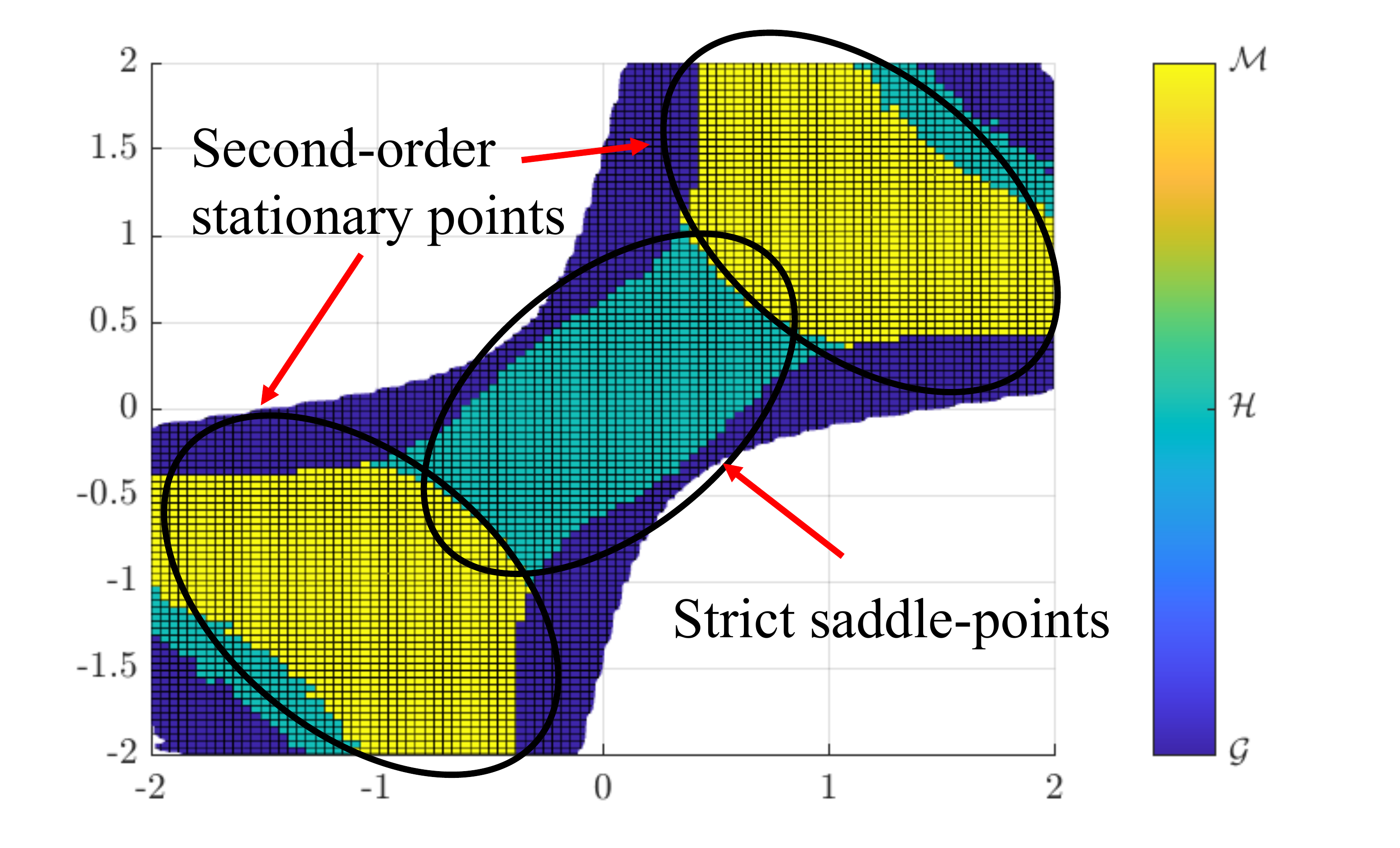}
  \caption{{\( 0.1 \)-stationary points.}}
\end{subfigure}
\begin{subfigure}[b]{.3\textwidth}
  \centering
  \includegraphics[width=\linewidth]{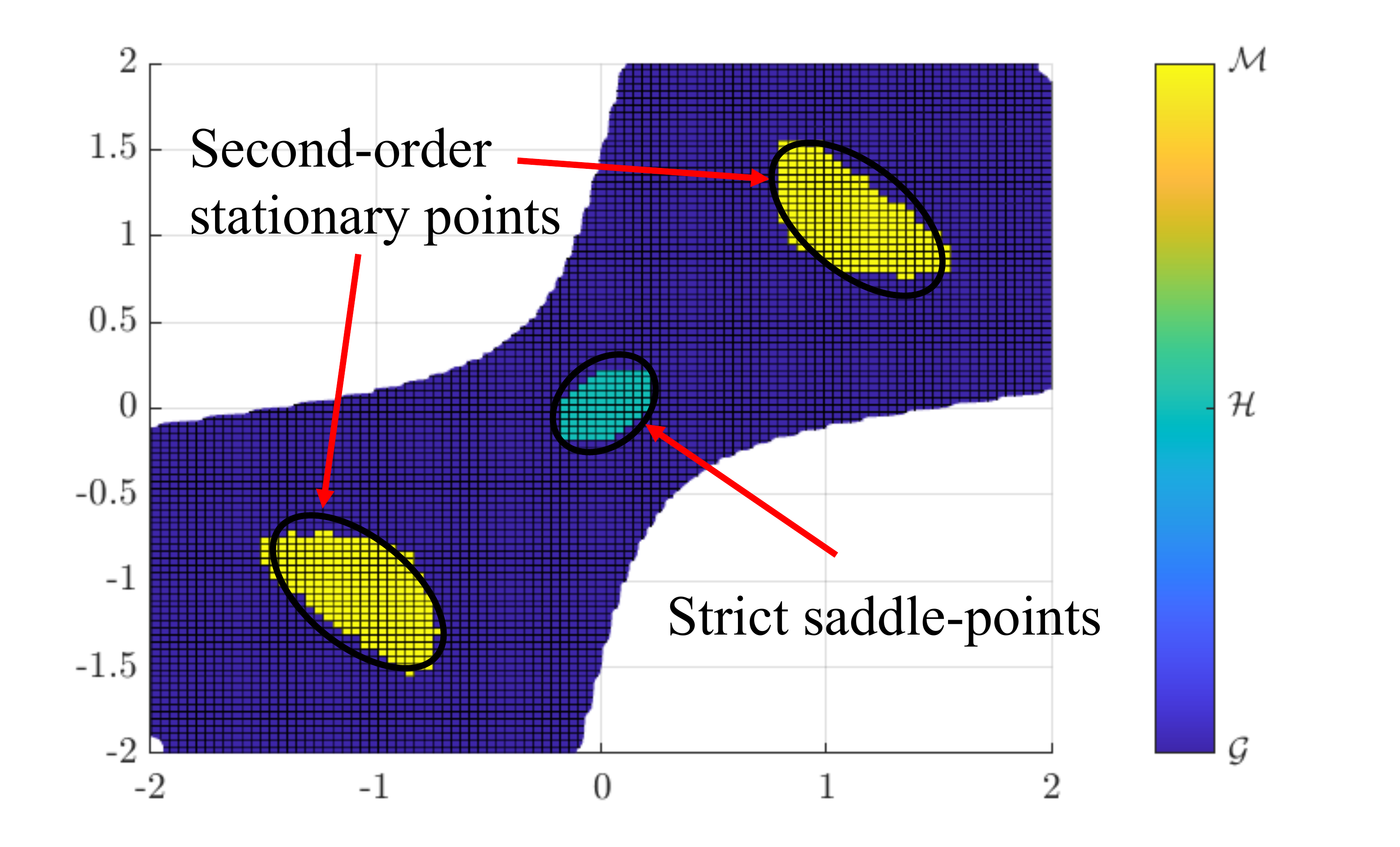}
  \caption{{\( 0.01 \)-stationary points.}}
\end{subfigure}
\caption{A visual representation of the space decomposition introduced in Definitions~\ref{def:first-order} through~\ref{def:strict-saddle} on a sample risk surface with two local minimizers and one saddle-point. The risk surface is depicted in Figure~(a). Points in space colored teal \( \mathcal{H} \) and yellow \( \mathcal{M} \) in Figures~(b) and~(c) are all \( 0.1 \) and \(0.01 \)-first-order stationary respectively according to Definition~\ref{def:first-order}. As such, first-order convergence guarantees can only guarantee that the algorithm does not return points in the complement of \( \mathcal{G} \), marked in purple, where the norm of the gradient is large. In contrast, we further decompose the set of first-order stationary points into the set of strict saddle-points (set \( \mathcal{H} \) in teal) and second-order stationary points (set \( \mathcal{M} \), yellow), establish descent for points in \( \mathcal{H} \) (teal), and conclude return of second-order stationary points in \( \mathcal{M} \) (yellow). Reduction of the step-size parameter \( \mu \) results in contraction of the set of approximate second-order stationary points around the true local minimizers as is observed from Figure~(b) to Figure~(c).}
\label{fig:space_decomp}
\end{figure}

This observation has motivated a number of works to pursue higher-order stationarity guarantees of local descent algorithms by means of second-order information~\cite{Nesterov06, Curtis17, Tripuraneni18}, intermediate searches for the negative curvature direction~\cite{Fang18, Allen18neon, Allen18natasha}, perturbations in the initialization~\cite{Lee16, Du17, Scutari18} or to the update direction~\cite{Gelfand91, Ge15, Jin17, Fang19, Jin19, HadiDaneshmand18, Swenson19, Vlaski19single, Vlaski19nonconvexP1, Vlaski19nonconvexP2}, both in the centralized and decentralized setting. Our focus in this manuscript will be on strategies that exploit the presence of perturbations in the update direction to escape from saddle-points. The motivation for this is two-fold. First, in large-scale and online learning problems, the evaluation of exact descent directions is generally infeasible, making the utilization of stochastic gradients, and hence the introduction of stochastic perturbations a necessity. Second, as we shall see, perturbations to the gradient direction can be shown to be sufficient to guarantee \emph{efficient} escape from saddle-points, meaning that  the escape-time can be bounded by quantities that scale favorably with problem dimensions and parameters, resulting in simple, yet effective solutions for escaping saddle-points and guaranteeing second-order stationarity without the need to significantly alter the operation of the algorithm.

\subsection{Stochastic Gradient Descent}
One popular first-order approach to pursuing a minimizer for problem~\eqref{eq:centralized_expected} can be obtained means of gradient descent, resulting in the recursion:
\begin{equation}\label{eq:gradient_descent}
  w_i = w_{i-1} - \mu \nabla J(w_{i-1})
\end{equation}
The limitation of this recursion lies in the fact that evaluation of the exact gradient of \( J(w_{i-1}) \) requires statistical information about the random variable \( \x \) in light of:
\begin{equation}
  \nabla J(w_{i-1}) \triangleq \nabla \left( \mathds{E}_{\boldsymbol{x}} Q(w; \boldsymbol{x} ) \right)
\end{equation}
The most common remedy for this challenge is to instead employ a stochastic approximations of the gradient \( \nabla J(w_{i-1}) \) based on realizations of the random variable \( \boldsymbol{x} \) available at time \( i \). We denote a general stochastic gradient approximation by \( \widehat{\nabla J}(\cdot) \) and iterate:
\begin{equation}\label{eq:sgd_general}
  \w_{i} = \w_{i-1} - \mu \widehat{\nabla J}(\w_{i-1})
\end{equation}
Observe that we now denote \( \w_{i} \) in bold font to emphasize the fact that, by utilizing a stochastic approximation \( \widehat{\nabla J}(\cdot) \) based on \emph{realizations} of the random variable \( \x \) in place of the true gradient \( {\nabla J}(\cdot) \) based on the \emph{distribution} of \( \x \), the resulting iterates \( \w_i \) will become stochastic themselves. We will leave the actual specification of the approximation \( \widehat{\nabla J}(\w_{i-1}) \) for the examples and describe performance guarantees under general approximations satisfying fairly general modeling conditions.

\subsection{Modeling Conditions}
We begin by introducing smoothness conditions on both the gradient and Hessian of the risk \( J(\cdot) \).
\begin{assumption}[\textbf{Lipschitz gradients}]\label{as:lipschitz}
  The gradient \( \nabla J(\cdot) \) is Lipschitz, namely, there exists \( \delta > 0 \) such that for any \( x,y \):
  \begin{equation}\label{eq:lipschitz}
    \|\nabla J(x) - \nabla J(y)\| \le \delta \|x-y\|
  \end{equation}
\end{assumption}\hfill\qed%
\begin{assumption}[\textbf{Lipschitz Hessians}]\label{as:lipschitz_hessians}
  The risk \( J(\cdot) \) is twice-differentiable and there exists \( \rho \ge 0 \) such that:
  \begin{equation}
    {\| \nabla^2 J(x) - \nabla^2 J(y) \|} \le \rho \|x - y\|
  \end{equation}\hfill\qed
\end{assumption}
Condition~\eqref{eq:lipschitz} appears commonly in the study of first-order optimality guarantees of (stochastic) gradient algorithms~\cite{Nesterov98, Bertsekas00, Sayed14}. The Lipschitz condition on the Hessian matrix is not necessary to establish performance bounds in the (strongly-)convex case or first-order stationarity, but can be used to more accurately quantify deviations around the minimizer in steady-state~\cite{Sayed14}, or to establish the escape from saddle-points~\cite{Ge15, Jin17, HadiDaneshmand18, Vlaski19nonconvexP1, Vlaski19nonconvexP2}. The second set of conditions below establishes bounds on the quality of the stochastic gradient approximation \( \widehat{\nabla J}(\cdot) \). We define the stochastic gradient noise process:
\begin{equation}\label{eq:gradient_noise_def}
  \s_i(\w_{i-1}) \triangleq {\nabla J}(\w_{i-1}) - \widehat{\nabla J}(\w_{i-1})
\end{equation}
\begin{assumption}[\textbf{Gradient noise process}]\label{as:gradientnoise}
  The gradient noise process~\eqref{eq:gradient_noise_def} is unbiased with a relative bound on its fourth-moment:
  \begin{align}
    \E \left\{ \s_{i}(\w_{i-1}) | \w_{i-1} \right\} &= 0 \label{eq:conditional_zero_mean}\\
    \E \left\{ \|\s_{i}(\w_{i-1})\|^4 | \w_{i-1} \right\} &\le \beta^4 {\left \| \nabla J(\w_{i-1}) \right\|}^4 + \sigma^4 \label{eq:gradientnoise_fourth}
  \end{align}
  for some non-negative constants \( \beta^4, \sigma^4 \).\hfill\qed
\end{assumption}
Relation~\eqref{eq:conditional_zero_mean} requires that the gradient approximation \( \widehat{\nabla J}(\cdot) \) be unbiased. Condition~\eqref{eq:gradientnoise_fourth} imposes a bound on the fourth moment of the gradient noise, but allows for this bound to grow with the norm of the gradient \( {\left \| \nabla J(\w_{i-1}) \right\|}^4 \). Note that, in light of Jensen's inequality and sub-additivity of the square root, condition~\eqref{eq:gradientnoise_fourth} implies and is slightly stronger than:
\begin{equation}
  \E \left\{ \|\s_{i}(\w_{i-1})\|^2 | \w_{i-1} \right\} \le \beta^2 {\left \| \nabla J(\w_{i-1}) \right\|}^2 + \sigma^2 \label{eq:gradientnoise}
\end{equation}
Condition~\eqref{eq:gradientnoise} is sufficient to establish limiting first-order stationarity~\cite{Bertsekas00}, while the fourth-moment condition~\eqref{eq:gradientnoise_fourth} will allow us to more carefully analyze the dynamics of~\eqref{eq:sgd_general} around first-order stationary points and establish escape from saddle-points, resulting in second-order guarantees. We also impose conditions on the covariance of the gradient noise.
\begin{assumption}[\textbf{Lipschitz covariances}]\label{as:lipschitz_covariance}
  The gradient noise process has a Lipschitz covariance matrix, i.e.,
  \begin{equation}\label{eq:covariance_def}
    R_{s}(\w_{i-1}) \triangleq \E \left \{ \s_{i}(\w_{i-1}) {\s_{i}(\w_{i-1})}^{\T} | \w_{i-1} \right \}
  \end{equation}
  satisfies
  \begin{equation}\label{eq:lipschitz_r}
    \| R_{s}(x) - R_{s}(y) \| \le \beta_R {\| x - y \|}^{\gamma}
  \end{equation}
  for all \( x, y \), some \( \beta_R \ge 0 \) and \( 0 < \gamma \le 4\).\hfill\qed
\end{assumption}
Note from the definition of the gradient noise covariance~\eqref{eq:covariance_def}, that the distribution of the gradient noise process is a function of the iterate \( \w_{i-1} \). This, of course, is natural since the gradient noise is defined in~\eqref{eq:gradient_noise_def} as the difference between the true and the approximate gradient \emph{at the current iterate}. The fact that the perturbations introduced into the stochastic recursion~\eqref{eq:sgd_general} are not necessarily identically distributed over time introduces challenges in the study of their cumulative effect. Thankfully, the gradient noise processes induced by most constructions for \( \widehat{\nabla J}(\cdot) \) and losses \( Q(\cdot, \cdot) \) of interest have a covariance with a Lipschitz-type property~\eqref{eq:lipschitz_r}. This condition ensures that the covariance \( R_{s}(\w_{i-1}) \) is sufficiently smooth over localized regions in space, resulting in essentially identically distributed gradient noise perturbations in the short-term and a tractable analysis. It has also been exploited to derive accurate steady-state performance expressions in the strongly-convex setting~\cite{Sayed14}.

In contrast to Assumption~\ref{as:gradientnoise}, which bounds the perturbations induced by employing stochastic gradient approximations \emph{from above}, we will also be imposing a lower bound on the stochastic gradient noise.
\begin{assumption}[\textbf{Gradient noise in strict saddle-points}]\label{as:noise_in_saddle}
  Suppose \( w \) is an approximate strict-saddle point following Definition~\ref{def:strict-saddle}. Introduce the eigendecomposition of the Hessian matrix as \( \nabla^2 J(w) = V \Lambda V^{\T} \) and partition:
  \begin{equation}
    {V} = \left[ \begin{array}{cc} {V}^{\ge0} & {V}^{< 0} \end{array} \right],
    \ \ {\Lambda} = \left[ \begin{array}{cc} {\Lambda}^{\ge0} & 0\\0 & {\Lambda}^{< 0} \end{array}\right]
  \end{equation}
  where \( {\Lambda}^{\ge0} \ge 0 \) and \( {\Lambda}^{< 0} < 0 \). Then, we assume that:
  \begin{equation}\label{eq:noise_in_saddle}
    \lambda_{\min}\left({\left({V}^{< 0}\right)}^{\T} {R}_{s}\left(w \right) {V}^{< 0} \right) \ge \sigma_{\ell}^2
  \end{equation}
  for some \( \sigma_{\ell}^2 > 0 \).\hfill\qed
\end{assumption}
\begin{figure}[!t]
	\centering
	\includegraphics[width=\columnwidth]{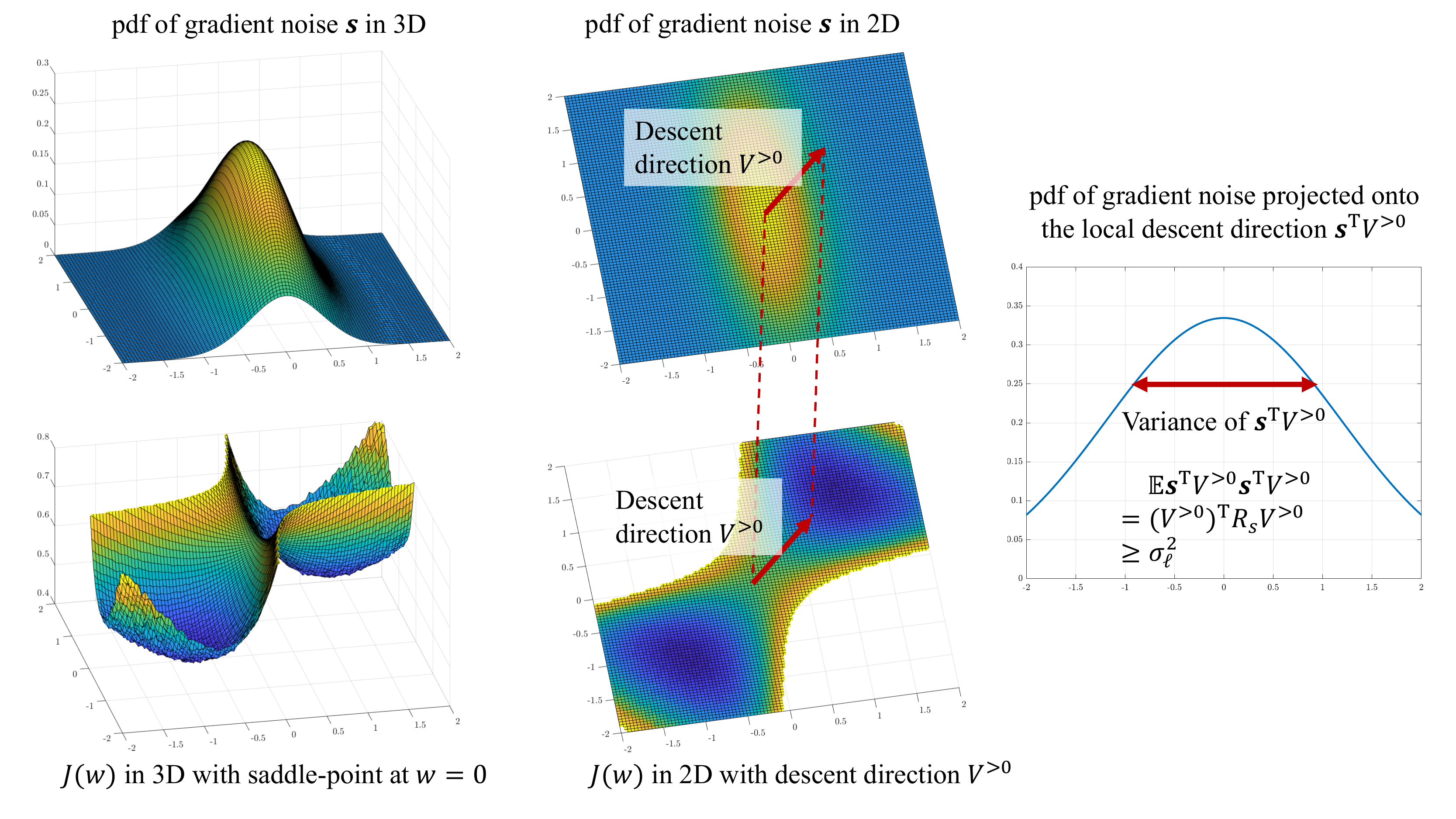}
	\caption{A visual illustration of Assumption~\ref{as:noise_in_saddle}, which imposes a lower bound on the alignment between gradient noise and the local descent direction. Examples of a probability density function of the gradient noise \( \s_i(\w_{i-1}) \) (top) and risk function \( J(w) \) (bottom) are shown in the left and middle columns, respectively. The risk \( J(w) \) exhibits a strict saddle-point at \( w = 0 \). The local descent direction, which corresponds to \( V^{>0} \) in~\eqref{eq:noise_in_saddle} is emphasized as a red arrow in the middle column. Assumption~\ref{as:noise_in_saddle} requires some alignment between the local descent direction \( V^{>0} \) of the risk (middle bottom) and the probability density function of the gradient noise process (middle top). The quantity \( {\left({V}^{< 0}\right)}^{\T} {R}_{s}\left(w \right) {V}^{< 0} \) in condition~\eqref{eq:noise_in_saddle} in this two-dimensional example corresponds precisely to the variance of the gradient noise after projecting \( \s_i(\w_{i-1}) \) onto the local descent direction \( V^{>0} \), shown in the right column.}\label{fig:alignment}
\end{figure}
{If we construct a local Taylor approximation around the strict saddle-points \( w \), we have:
\begin{align}
  J(w + \Delta w) \approx&\: J(w) + \nabla J(w)^{\T}\Delta w + \Delta w^{\T} \nabla^2 J(w) \Delta w \notag \\
  \approx&\: J(w) + \Delta w^{\T} \nabla^2 J(w) \Delta w
\end{align}
since at strict saddle-points \( \nabla J(w) \approx 0 \) and, hence, the linear term vanishes. For every \( \Delta w  \) in the range of \( V^{<0} \), i.e., \( \Delta w \triangleq V^{<0} x \) , we then have \( \Delta w^{\T} \nabla^2 J(w) \Delta w = x^{\T} {\left(V^{<0}\right)}^{\T} \nabla^2 J(w) V^{<0} x < 0 \) by definition of \( V^{<0} \), and hence \( J(w + \Delta w) < J(w) \). We conclude that the space spanned by \( V^{<0} \) corresponds to the local descent directions around the strict saddle-point \( w \). Hence}, condition~\eqref{eq:noise_in_saddle} imposes a lower bound on the gradient noise component in the local descent direction (spanned by \( {V}^{< 0} \)) in the vicinity of saddle-points{. It is} a notable deviation from the assumptions typically imposed in the \emph{convex} setting. While assumptions~\ref{as:lipschitz}--\ref{as:lipschitz_covariance} are for example all leveraged in deriving steady-state performance expressions in~\cite{Sayed14} under an additional strong-convexity condition, assumption~\ref{as:noise_in_saddle} is unique to the study of the behavior of stochastic gradient-type algorithms in the vicinity of saddle-points~\cite{HadiDaneshmand18, Vlaski19nonconvexP1, Vlaski19nonconvexP2} in nonconvex optimization. It may be particularly surprising since the presence of perturbations in the dynamics of gradient-type algorithms are generally understood to be negative side-effects of the utilization of stochastic gradient approximations and result in deterioration of performance, which is generally true for (strongly) convex objectives. {When generalizing to nonconvex objectives, as} recent analysis has shown~\cite{Ge15, Jin17, HadiDaneshmand18, Vlaski19nonconvexP1, Vlaski19nonconvexP2}, the persistent presence of gradient perturbations allows the algorithm to efficiently escape from saddle-points, which are unstable to gradient perturbations, and arrive at local minima, which tend to be more stable to the same types of perturbations. In this sense, condition~\eqref{eq:noise_in_saddle} allows the algorithm to distinguish stable local minima from unstable saddle-points, both of which are first-order stationary points.

As we will see in the examples in the sequel, and the following Section~\ref{sec:federated_learning}, the formulation~\eqref{eq:sgd_general} under the modeling conditions~\ref{as:lipschitz}--\ref{as:noise_in_saddle} is sufficiently general to capture a plethora of first-order stochastic algorithms for the minimization of~\eqref{eq:define_j}.
\begin{example}[\textbf{Stochastic gradient descent}]\label{ex:sgd}
  Suppose we have access to a realization of the data \( \x_i \) at time \( i \). We can construct a stochastic gradient approximation as:
  \begin{equation}\label{eq:sgd_gradient}
    \widehat{\nabla J}^{\mathrm{SGD}}(\w_{i-1}) \triangleq \nabla Q(\w_{i-1}; \x_i)
  \end{equation}
  Then, condition~\eqref{eq:conditional_zero_mean} follows immediately by definition of~\eqref{eq:sgd_gradient}, while~\eqref{eq:gradientnoise_fourth} can be verified for a number of choices of the loss function \( Q(w; \boldsymbol{x}) \) and data distributions of \( \boldsymbol{x} \). We shall denote the resulting constants:
  \begin{equation}\label{eq:sgd_bounds}
    \E \left\{ \|\s_i^{\mathrm{SGD}}(\w_{i-1})\|^4 | \w_{i-1} \right\} \le \beta_{\mathrm{SGD}}^4 {\left \| \nabla J(\w_{i-1}) \right\|}^4 + \sigma_{\mathrm{SGD}}^4
  \end{equation}\hfill\qed%
\end{example}
\begin{example}[\textbf{Mini-batch stochastic gradient descent}]
  Suppose we instead have access to a collection of \( B \) independent samples \( \left\{ \x_{b, i} \right\}_{b=1}^B \) at time \( i\) and the computational capacity to compute \( B \) gradient operations at every iteration. We can then construct the mini-batch gradient approximation:
  \begin{equation}\label{eq:b_sgd_gradient}
    \widehat{\nabla J}^{\mathrm{B-SGD}}(\w_{i-1}) \triangleq \frac{1}{B} \sum_{b=1}^B \nabla Q(\w_{i-1}; \x_{b, i})
  \end{equation}
  It again follows that \( \widehat{\nabla J}^{\mathrm{B-SGD}}(\w_{i-1}) \) satisfies~\eqref{eq:conditional_zero_mean}. For the fourth-order moment can verify by induction over \( B \) that:
  \begin{align}\label{eq:b_squared}
    \E \left\{ \left \|\s_i^{\mathrm{B-SGD}}(\w_{i-1}) \right \|^4 | \w_{i-1} \right\} \le C_B \left( \frac{\beta_{\mathrm{SGD}}^4}{B^2} {\left \| \nabla J(\w_{i-1}) \right\|}^4 + \frac{\sigma_{\mathrm{SGD}}^4}{B^2} \right)
  \end{align}
  in terms of the constants \( \beta_{\mathrm{SGD}}^4 \) and \( \sigma_{\mathrm{SGD}}^4 \) of the single-element stochastic gradient algorithm in example~\ref{ex:sgd}, as well as the constant:
  \begin{equation}
    C_B \triangleq 3 - \frac{2}{B} \le 3
  \end{equation}
  We observe a \( B^2 \)-fold decrease in the mean-fourth moment, which implies a \( B \)-fold reduction in the second-order moment and complies with our intuition about variance reduction by averaging. For the gradient noise covariance we have:
  \begin{equation}
    R_s^{\mathrm{B-SGD}}(\w_{i-1}) = \frac{1}{B} R_s^{\mathrm{SGD}}(\w_{i-1})
  \end{equation}\hfill\qed%
\end{example}
\begin{example}[\textbf{Perturbed stochastic gradient descent}]
  In the absence of prior knowledge that there is a gradient noise component in the descent direction for every strict saddle-point (Assumption~\ref{as:noise_in_saddle}), one can always guarantee condition~\eqref{eq:noise_in_saddle} to hold by adding a small perturbation term \( \boldsymbol{v}_i \) with positive-definite covariance matrix \( R_{v} \triangleq \E \boldsymbol{v} \boldsymbol{v}^{\T} > 0 \) as done in~\cite{Ge15, Jin19} to construct:
  \begin{equation}
    \widehat{\nabla J}^{\mathrm{P-SGD}}(\w_{i-1}) \triangleq \widehat{\nabla J}^{\mathrm{SGD}}(\w_{i-1}) + \boldsymbol{v}_i = \nabla Q(\w_{i-1}; \x_i) + \boldsymbol{v}_i
  \end{equation}
  For the gradient noise covariance we then have:
  \begin{equation}
    R_s^{\mathrm{P-SGD}}(\w_{i-1}) = R_s^{\mathrm{SGD}}(\w_{i-1}) + R_v > 0
  \end{equation}
  and hence Assumption~\ref{as:noise_in_saddle} is guaranteed to hold. More elaborate constructions, such as only adding an additional perturbation when the iterate \( \w_{i-1} \) is suspected to be near a first-order stationary point (as done in~\cite{Jin17}) are also possible.\hfill\qed%
\end{example}

\subsection{Second-Order Guarantee}
Due to space limitations, we will only outline the main results that lead to a second-order guarantee for the stochastic approximation algorithm~\eqref{eq:sgd_general} and refer the reader to~\cite{Vlaski19single} for a thorough derivation of the result. We begin by formalizing the space decomposition into first and second-order stationary points as well as strict saddle-points.
\begin{definition}[Sets]\label{DEF:SETS}
  To simplify the notation in the sequel, we introduce the following sets:
  \begin{align}
    \mathcal{G} &\triangleq \left \{ w : {\left \| \nabla J(w) \right \|}^2 \ge \mu \frac{c_2}{c_1}\left(1+ \frac{1}{\pi}\right) \right \} \label{eq:define_g}\\
    \mathcal{G}^C &\triangleq \left \{ w : {\left \| \nabla J(w) \right \|}^2 < \mu \frac{c_2}{c_1} \left(1+\frac{1}{\pi} \right)\right \} \\
    \mathcal{H} &\triangleq \left \{ w : w \in \mathcal{G}^C, \lambda_{\min}\left( \nabla^2 J(w) \right) \le -\tau \right \} \label{eq:define_h}\\
    \mathcal{M} &\triangleq \left \{ w : w \in \mathcal{G}^C, \lambda_{\min}\left( \nabla^2 J(w) \right) > -\tau \right \} \label{eq:define_m}
  \end{align}
  where \( \tau \) is a small positive parameter, \( c_1 \) and \( c_2 \) are constants:
  \begin{align}
		c_1 &\triangleq  1 - \mu \frac{\delta}{2} \left( 1+\beta^2 \right) = O(1) \label{eq:define_c1}\\
		c_2 &\triangleq \frac{\delta}{2} \sigma^2 = O(1) \label{eq:define_c2}
	\end{align}
  and \( 0 < \pi < 1 \) is a parameter to be chosen. Note that \( \mathcal{G}^C = \mathcal{H} \cup \mathcal{M} \). For brevity, we also define the probabilities \( \pi^{\mathcal{G}}_i \triangleq \mathrm{Pr}\left \{ \w_{i} \in \mathcal{G} \right \} \), \(\pi^{\mathcal{H}}_i \triangleq \mathrm{Pr}\left \{ \w_{i} \in \mathcal{H} \right \} \) and \(\pi^{\mathcal{M}}_i \triangleq \mathrm{Pr}\left \{ \w_{ i} \in \mathcal{M} \right \}\). Then, for all \( i \), we have \( \pi^{\mathcal{G}}_i + \pi^{\mathcal{H}}_i + \pi^{\mathcal{M}}_i = 1 \).\hfill\qed%
\end{definition}
The set \( \mathcal{G}^C \) formalizes the set of \( O(\mu) \)-first-order stationary points in Definition~\ref{def:first-order} by setting the constant multiplying the step-size \( \mu \) to \( \frac{c_2}{c_1}\left(1+ \frac{1}{\pi}\right) \) where \( c_1, c_2 \) are problem-dependent constants. The set \( \mathcal{M} \) then corresponds to the set of second-order stationary points in Definition~\ref{def:second-order} while \( \mathcal{H} \) denotes the set of strict saddle-points in Definition~\ref{def:strict-saddle}. For a visualization, we refer the reader back to Fig.~\ref{fig:space_decomp}.

Points in both \( \mathcal{G} \) and \( \mathcal{H} \) are ``undesirable'' limiting points in the sense that they have local directions of descent. Our objective is to show that for iterates within both sets, algorithm~\eqref{eq:sgd_general} will continue to descend along the risk~\eqref{eq:define_j} by taking local gradient steps. The two sets \( \mathcal{G} \) and \( \mathcal{H} \) are distinguished by the fact that for points in \( \mathcal{G} \), the gradient norm \( {\left \| \nabla J(w) \right \|}^2 \) is large enough for a single (stochastic) gradient step to be sufficient to guarantee descent in expectation. Points in \( \mathcal{H} \) (i.e., strict saddle-points) on the other hand are more challenging since the gradient norm is so small that a single gradient step is no longer sufficient to guarantee descent.

\begin{theorem}[\textbf{Descent in the large-gradient regime~\cite{Vlaski19single}}]\label{LEM:DESCENT_RELATION}
  For sufficiently small step-sizes:
  \begin{equation}
    \mu \le \frac{2}{\delta\left(1+\beta^2\right)}
  \end{equation}
  and when the gradient at \( \w_i \) is sufficiently large, i.e., \( \w_i \in \mathcal{G} \), the stochastic gradient recursion~\eqref{eq:sgd_general} yields descent in expectation in one iteration, namely,
  \begin{align}\label{eq:one_step_descent}
    \E \left \{ J(\w_{i+1}) | \w_i \in \mathcal{G} \right \} \le \E \left \{ J(\w_{i}) | \w_i \in \mathcal{G} \right \} - \mu^2 \frac{c_2}{\pi}
  \end{align}
  {We also establish the following technical result, which bounds the negative effect of the gradient noise close to local minima \( w \in \mathcal{M} \)}:
  \begin{align}
    \E \left \{ J(\w_{i+1}) | \w_i \in \mathcal{M} \right \} \le \E \left \{ J(\w_{i}) | \w_i \in \mathcal{M} \right \} + \mu^2 {c_2}
  \end{align}\hfill\qed%
\end{theorem}
In the vicinity of strict saddle-points \( \mathcal{H} \), a more detailed analysis is necessary. Here, it is not the gradient step that ensures descent, but rather the cumulative effect of the gradient noise perturbations to the gradient update. The definition of a strict saddle-point~\eqref{eq:define_h} ensures that there is a direction of negative curvature in the local risk surface, while Assumption~\ref{as:noise_in_saddle} guarantees that with some probability the iterate \( \w_i \) is perturbed towards the descent direction. Together, these conditions allow the algorithm to escape along the descent direction with high probability in a finite number of iterations. This intuition is formalized by constructing a local short-term model based on a local quadratic approximation of the risk surface with identically distributed gradient perturbations and exploiting the smoothness conditions~\ref{as:lipschitz_hessians} and~\ref{as:lipschitz_covariance} to bound the approximation error~\cite[Lemma 3]{Vlaski19single}.
\begin{theorem}[\textbf{Descent through strict saddle-points~\cite{Vlaski19single}}]\label{TH:DESCENT_THROUGH_SADDLE_POINTS}
  Beginning at a strict saddle-point \( \w_i \in \mathcal{H} \) and iterating for \( i^s \) iterations after \( i \) with
  \begin{align}\label{eq:escape_time}
    i^{s} = \frac{\log\left( 2 M  \frac{\sigma^2}{\sigma_{\ell}^2} + 1 + O(\mu) \right)}{\log({1 + 2\mu\tau})} \le O\left(\frac{1}{\mu \tau} \right)
  \end{align}
  guarantees
  \begin{align}
    \E \left \{ J(\w_{i+i^s}) | \w_{i} \in \mathcal{H} \right \} \le \E \left \{ J(\w_{i}) | \w_{i} \in \mathcal{H} \right \} - \frac{\mu}{2} M \sigma^2 + o(\mu)
  \end{align}\hfill\qed%
\end{theorem}
Theorem~\ref{TH:DESCENT_THROUGH_SADDLE_POINTS} ensures that, even when the norm of the gradient is too small to carry sufficient information about the descent direction, the gradient noise along with the negative local curvature of the risk surface around strict saddle-points is sufficient to guarantee descent in \( i^s \) iterations, where the escape-time scales favorably with problem parameters. For example, the escape time scales logarithmically with the problem dimension \( M \), implying that we can expect fast evasion of saddle-points even in high dimensions. Having established descent both in the large-gradient regime and strict-saddle point regime, we can combine the results to conclude eventual second-order stationarity.
\begin{theorem}[\textbf{Second-order guarantee for stochastic gradient descent \cite{Vlaski19single}}]\label{TH:FINAL_THEOREM}
  Suppose \( J(w) \ge J^o \). Then, for sufficiently small step-sizes \( \mu \), we have with probability \( 1 - \pi \), that \( \w_{i^o} \in \mathcal{M} \), i.e., \( \| \nabla J(\w_{i^o}) \|^2 \le O(\mu) \) and \( \lambda_{\min}\left( \nabla^2 J(\w_{i^o}) \right) \ge -\tau \) in at most \( i^o \) iterations, where
  \begin{align}
    i^o \le \frac{\left( J(w_{0}) - J^o \right)}{\mu^2 c_2 \pi} i^s,
  \end{align}
  the quantity \( J(w_{0}) - J^o \) denotes the sub-optimality at the initialization \( w_0 \) and \( i^s \) denotes the escape time from Theorem~\ref{TH:DESCENT_THROUGH_SADDLE_POINTS}.
\end{theorem}

\section{Federated Learning}\label{sec:federated_learning}
In many large-scale applications, data is not available at a central processor, but is instead collected and processed at distributed locations. In this section, we consider a multi-agent setting with a collection of \( K \) agents and a central node for parameter aggregation. We associate with each agent its own risk~\eqref{eq:centralized_expected}, indexed by \( k \):
\begin{equation}\label{eq:federated_local_problem}
  J_k(w) \triangleq \E_{\x_k} Q_k(w; \x_k)
\end{equation}
and would like to pursue the minimizer of:
\begin{equation}\label{eq:federated_problem}
  J(w) \triangleq \sum_{k=1}^K p_k J_k(w)
\end{equation}
where \( p_k > 0 \) denote positive weights, normalized to add up to one without loss of generality, i.e., \( \sum_{k=1}^K p_k = 1 \). The problem of distributed minimization of~\eqref{eq:federated_problem} in the presence of a centralized processor, but without the aggregation of raw data \( \x_{k, i} \) can be achieved using primal and primal-dual approaches~\cite{Zinkevich10, Duchi09}. More recently, federated learning has emerged as a framework for the solution of~\eqref{eq:federated_problem} under considerations of asynchrony, heterogeneity, communication and computational restrictions and privacy concerns as they are encountered in practical applications~\cite{mcmahan16}. We show in the sequel that a version of the federated averaging algorithm~\cite{mcmahan16} can be interpreted as the construction of a particular choice for the stochastic gradient approximation \( \widehat{\nabla J}(\cdot) \), and hence second-order guarantees can be obtained directly by specializing the results from Section~\ref{sec:centralized}. As a baseline, consider the true gradient update to~\eqref{eq:federated_problem}, which takes the form:
\begin{equation}\label{eq:federated_gradient_descent}
  w_{i} = w_{i-1} - \mu \nabla J(w_{i-1}) = w_{i-1} - \mu \sum_{k=1}^K p_k \nabla J_k(w_{i-1})
\end{equation}
Just like its single-agent counter-part~\eqref{eq:gradient_descent}, recursion~\eqref{eq:federated_gradient_descent} has the drawback of requiring statistical information about \( \x_k \) to evaluate the expectations in~\eqref{eq:federated_local_problem}. Additionally,~\eqref{eq:federated_gradient_descent} requires full and synchronous participation of all agents \( k \) at every iteration by providing (or approximating) the local gradient \( \nabla J_k(w_{i-1}) \). The former issue can be addressed by employing stochastic gradient approximations based on realizations of data from the distribution of \( \x_{k, i} \), while the latter issue can be relaxed by allowing for partial participation of agents. To this end, at every iteration \( i \), we sample \( L \) agent-indices without replacement from the set \( \{ 1, \ldots, K \} \) to form \( \mathcal{L} \). We introduce the participation indicator function:
\begin{equation}
  \mathds{1}_{k, i} = \begin{cases} 1, \ \mathrm{if}\ k \in \mathcal{L}\ \mathrm{at}\ \mathrm{iteration}\ i, \\ 0, \ \mathrm{otherwise.} \end{cases}
\end{equation}
Then, at every iteration, the global model \( \w_{i-1} \) is broadcast to participating agents, which collect local data \( \left\{ \x_{k, i, b} \right\}_{b=1}^B \) and perform the update:
\begin{equation}\label{eq:federated_local_update}
  \w_{k, i} = \w_{i-1} - \mu K \mathds{1}_{k, i}  \frac{p_k}{B} \sum_{b=1}^B \nabla Q_k(\w_{i-1}; \x_{k, i, b})
\end{equation}
The central processor can then aggregate the intermediate estimates from the participating agents and compute:
\begin{equation}\label{eq:federated_combination}
  \w_i = \frac{1}{L} \sum_{k=1}^K \mathds{1}_{k, i} \w_{k, i}
\end{equation}
Due to the presence of the indicator function \( \mathds{1}_{k, i} \), the aggregation step~\eqref{eq:federated_combination} only requires exchanges with participating agents. Steps~\eqref{eq:federated_local_update} and~\eqref{eq:federated_combination} can be combined into:
\begin{equation}\label{eq:federated_together}
  \w_{i} = \w_{i-1} - \mu \frac{K}{L} \sum_{k=1}^K \mathds{1}_{k, i} \frac{p_k}{B} \sum_{b=1}^B \nabla Q_k(\w_{i-1}; \x_{k, b, i})
\end{equation}
We argue in the sequel that the approximation:
\begin{equation}
  \widehat{\nabla J}(\w_{i-1}) \triangleq \frac{K}{ L} \sum_{k=1}^K \mathds{1}_{k, i}\frac{p_k}{B} \sum_{b=1}^B \nabla Q_k(\w_{i-1}; \x_{k, b, i})
\end{equation}
can be viewed as an instance of the stochastic approximation introduced in Section~\ref{sec:centralized} and hence the results from the single-agent analysis apply. In addition to assuming each \( J_k(\cdot) \) satisfies Assumptions~\ref{as:lipschitz}--\ref{as:noise_in_saddle}, we will impose the following bound on the agent heterogeneity~\cite{Swenson19, Vlaski19nonconvexP1}.
\begin{assumption}[\textbf{Bounded gradient disagreement}]\label{as:bounded}
  For each pair of agents \( k \) and \( \ell \), the gradient disagreement is bounded, namely, for any \( x \in \mathds{R}^{M} \):
  \begin{equation}\label{eq:bounded}
    \|\nabla J_k(x) - \nabla J_{\ell}(x)\| \le G
  \end{equation}\hfill\qed%
\end{assumption}
Relation~\eqref{eq:bounded} ensures that the disagreement on the local descent direction for any pair of agents is bounded, and is weaker than the more common assumption of uniformly bounded absolute gradients. From Jensen's inequality, we similarly bound the deviation from the aggregate gradient:
\begin{align}
  &\: \left \|\nabla J_k(x) - \nabla J(x) \right \| =\left \| \sum_{\ell=1}^N p_{\ell}\left( \nabla J_k(x) - \nabla J_{\ell}(x) \right)\right\| \notag \\
  \le&\: \sum_{\ell=1}^N p_{\ell} \left\| \nabla J_k(x) - \nabla J_{\ell}(x)\right)\| \le G
\end{align}

\begin{example}[Federated averaging as a centralized stochastic gradient approximation]\label{ex:federated}
  We define the local gradient approximation:
  \begin{align}
    \widehat{\nabla J}_k(\w_{i-1}) \triangleq&\: \frac{K}{L} \frac{\mathds{1}_{k, i}}{B} \sum_{b=1}^B \nabla Q_k(\w_{i-1}; \x_{k, b, i})\\
    \s_{k, i}(\w_{i-1}) \triangleq&\: \widehat{\nabla J}_k(\w_{i-1}) - {\nabla J}_k(\w_{i-1})
  \end{align}
  We then have:
  \begin{align}\label{eq:locally_unbiased}
    \E \left\{ \widehat{\nabla J}_k(\w_{i-1}) | \w_{i-1} \right\} \triangleq&\: \frac{K}{L} \frac{\E \left\{ \mathds{1}_{k, i} \right\}}{B} \sum_{b=1}^B \mathds{E} \left\{ \nabla Q_k(\w_{i-1}; \x_{k, b, i}) | \w_{i-1} \right\} \notag \\
    =&\: \frac{K}{L} \frac{L}{K} \frac{1}{B} \sum_{b=1}^B {\nabla J_k}(\w_{i-1}) = \nabla J_k(\w_{i-1})
  \end{align}
  where we used the fact that \(\E \left\{ \mathds{1}_{k, i} \right\} = \frac{L}{K} \). For the aggregate risk we then find:
  \begin{align}
    \E \left\{ \widehat{\nabla J}(\w_{i-1}) | \w_{i-1} \right\} = \sum_{k=1}^K p_k \E \left\{ \widehat{\nabla J}_k(\w_{i-1}) | \w_{i-1} \right\} \stackrel{\eqref{eq:locally_unbiased}}{=} {\nabla J}(\w_{i-1})
  \end{align}
  For the fourth-order moment we have:
  \begin{align}
    \E \left\{ {\| \s_{i}(\w_{i-1})\|}^4 | \w_{i-1} \right\}=&\: \E \left\{ {\left \| \sum_{k=1}^K p_k \s_{k, i}(\w_{i-1}) \right \|}^4 | \w_{i-1} \right\} \notag \\
    \stackrel{(a)}{\le}&\: \sum_{k=1}^K p_k \E \left\{ {\left \| \s_{k, i}(\w_{i-1}) \right \|}^4 | \w_{i-1} \right\}\label{eq:aggregate_in_terms_of_local}
  \end{align}
  where \( (a) \) follows from the convexity of \( \|\cdot\|^4 \) and Jensen's inequality. For the local gradient noise terms we have:
  {\begin{align}
    &\: \E \left\{ {\left \| \s_{k, i}(\w_{i-1}) \right \|}^4 | \w_{i-1} \right\} \notag \\
    =&\: \E \left\{ {\left \| \frac{K}{L} \frac{\mathds{1}_{k, i}}{B} \sum_{b=1}^B \nabla Q_k(\w_{i-1}; \x_{k, b, i}) - {\nabla J}_k(\w_{i-1}) \right \|}^4 | \w_{i-1} \right\} \notag \\
    =&\: \E \Bigg\{ \Bigg \| \frac{K}{L} \mathds{1}_{k, i} \left( \frac{1}{B} \sum_{b=1}^B \nabla Q_k(\w_{i-1}; \x_{k, b, i}) - {\nabla J}_k(\w_{i-1}) \right) \notag \\
    &\: \ \ \ \ \ \ \ + \left( \frac{K}{L} \mathds{1}_{k, i} -1 \right) {\nabla J}_k(\w_{i-1})   \Bigg \|^4 | \w_{i-1} \Bigg\} \notag \\
    \stackrel{(a)}{\le}&\: 8\frac{K^4}{L^4} \E \left\{ \mathds{1}_{k, i}^4 \right\} \cdot \E \left\{ {\left \| \frac{1}{B} \sum_{b=1}^B \nabla Q_k(\w_{i-1}; \x_{k, b, i}) - {\nabla J}_k(\w_{i-1}) \right \|}^4 | \w_{i-1} \right\} \notag \\
    &\: + 8\E \left\|\frac{K}{L} \mathds{1}_{k, i} -1\right\|^4 \left\|  {\nabla J}_k(\w_{i-1})   \right \|^4 \notag \\
    \stackrel{(b)}{=}&\: 8\frac{K^4}{L^4} \frac{L}{K} \cdot \E \left\{ {\left \| \frac{1}{B} \sum_{b=1}^B \nabla Q_k(\w_{i-1}; \x_{k, b, i}) - {\nabla J}_k(\w_{i-1}) \right \|}^4 | \w_{i-1} \right\} \notag \\
    &\: + 8 \left( \frac{L}{K} \frac{K-L}{L} + \frac{K-L}{K} \right) \left\|  {\nabla J}_k(\w_{i-1})   \right \|^4 \notag \\
    \stackrel{\eqref{eq:b_squared}}{\le}&\: \left( 8 \frac{K^3}{L^3} C_B \frac{\beta_{\mathrm{SGD}}^4}{B^2} + 16 \frac{K-L}{K} \right) {\left \| \nabla J_k(\w_{i-1}) \right\|}^4 + 8 \frac{K^3}{L^3} C_B \frac{\sigma_{\mathrm{SGD}}^4}{B^2} \notag \\
    =&\: \left( 8 \frac{K^3}{L^3} C_B \frac{\beta_{\mathrm{SGD}}^4}{B^2} + 16 \frac{K-L}{K} \right) {\left \| \nabla J_k(\w_{i-1}) - \nabla J(\w_{i-1}) + \nabla J(\w_{i-1})\right\|}^4 \notag \\
    &\: + 8 \frac{K^3}{L^3} C_B \frac{\sigma_{\mathrm{SGD}}^4}{B^2} \notag \\
    \stackrel{(c)}{\le}&\: \left( 64 \frac{K^3}{L^3} C_B \frac{\beta_{\mathrm{SGD}}^4}{B^2} + 128 \frac{K-L}{K} \right) \left( {\left \| \nabla J(\w_{i-1})\right\|}^4 + G^4 \right) + 8 \frac{K^3}{L^3} C_B \frac{\sigma_{\mathrm{SGD}}^4}{B^2}\notag \\
    \stackrel{(d)}{\le}&\: \beta_{\mathrm{Fed}}^4 {\left \| \nabla J(\w_{i-1}) \right\|}^4 + \sigma_{\mathrm{Fed}}^4
  \end{align}
  where \( (a) \) follows from Jensen's inequality \( \|a+b\|^4 \le 8\|a\|^4 + 8 \|b\|^4 \), \( (b) \) follows from \( \mathrm{Pr}\left\{ \mathds{1}_{k, i} = 1 \right\} = \frac{L}{K} \), \( (c) \) again follows from Jensen's inequality along with Assumption~\ref{as:bounded} and in \( (d) \) we defined:
  \begin{align}
    \beta_{\mathrm{Fed}}^4 \triangleq&\: 64 \frac{K^3}{L^3} C_B \frac{\beta_{\mathrm{SGD}}^4}{B^2} + 128 \frac{K-L}{K} \label{eq:federated_beta}\\
    \sigma_{\mathrm{Fed}}^4 \triangleq&\: \left( 64 \frac{K^3}{L^3} C_B \frac{\beta_{\mathrm{SGD}}^4}{B^2} + 128 \frac{K-L}{K} \right) G^4 + 8 \frac{K^3}{L^3} C_B \frac{\sigma_{\mathrm{SGD}}^4}{B^2} \label{eq:federated_sigma}
  \end{align}}
  For the aggregate gradient noise term we then have similarly from~\eqref{eq:aggregate_in_terms_of_local}:
  \begin{align}
    \E \left\{ {\left \| \s_{k, i}(\w_{i-1}) \right \|}^4 | \w_{i-1} \right\} \le \beta_{\mathrm{Fed}}^4 {\left \| \nabla J(\w_{i-1}) \right\|}^4 + \sigma_{\mathrm{Fed}}^4
  \end{align}
  and hence the federated learning algorithm~\eqref{eq:federated_together} satisfies Assumption~\ref{as:gradientnoise}.\hfill\qed%
\end{example}

\subsection{Second-Order Guarantees for Federated Averaging}
Having established in Example~\ref{ex:federated} that the federated averaging algorithm~\eqref{eq:federated_together} satisfies Assumption~\ref{as:gradientnoise} with constants~\eqref{eq:federated_beta}--\eqref{eq:federated_sigma}, we can specialize Theorem~\ref{TH:FINAL_THEOREM} to recover second-order guarantees for~\eqref{eq:federated_together}.
\begin{corollary}[\textbf{Second-order guarantee for federated averaging}]\label{TH:FEDERATED_GUARANTEE}
  Suppose \( J(w) \ge J^o \). Then, for sufficiently small step-sizes \( \mu \), the federated averaging algorithm~\eqref{eq:federated_together} with probability \( 1 - \pi \) generates a point \( \w_{i^o} \in \mathcal{M} \) with:
  \begin{align}
    \| \nabla J(\w_{i^o}) \|^2 \le O\left( \mu {\sigma_{\mathrm{Fed}}^2} \right)
  \end{align}
  and \( \lambda_{\min}\left( \nabla^2 J(\w_{i^o}) \right) \ge -\tau \) in at most \( i^o \) iterations, where
  \begin{align}
    i^o \le \frac{\left( J(w_{0}) - J^o \right)}{\mu^2 c_2 \pi} i^s,
  \end{align}
  {the quantity \( J(w_{0}) - J^o \) describes the initial sub-optimality,} and \( i^s \) denotes the escape time from Theorem~\ref{TH:DESCENT_THROUGH_SADDLE_POINTS}:
  \begin{equation}
    i^{s} = \frac{\log\left( 2 M  \frac{\sigma_{\mathrm{Fed}}^2}{\sigma_{\mathrm{Fed},\ell}^2} + 1 + O(\mu) \right)}{\log({1 + 2\mu\tau})} \le O\left(\frac{1}{\mu \tau} \right)
  \end{equation}\hfill\qed%
\end{corollary}
The constant \( {\sigma_{\mathrm{Fed}}^2} \) is the dominant constant determining the level of accuracy guaranteed by the result in Corollary~\ref{TH:FEDERATED_GUARANTEE}. Its expression in~\eqref{eq:federated_sigma} quantifies the dependence on the various federated learning parameters such as the participation rate \( \frac{L}{K} \) and the local mini-batch sizes \( B \).

\section{Decentralized Learning}\label{sec:decentralized_learning}
While fusion-center based approaches, such as~\eqref{eq:federated_together}, are an effective approach to learning from distributed data sources \( \x_k \) without the need to exchange raw data, and instead relying solely on the exchange of local models \( \w_{k, i} \), they have the drawback of nevertheless requiring some form of central aggregation. In this section, we relax this requirement. We continue to consider a collection of \( K \) agents, and continue to pursue solutions to~\eqref{eq:federated_problem}, repeated here for reference:
\begin{equation}\label{eq:decentralized_problem}
  J(w) \triangleq \sum_{k=1}^K p_k J_k(w)
\end{equation}
In contrast to the federated learning framework, which allows for the central aggregation of (a subset of) intermediate parameter estimates, we now consider the agents to be connected via a graph topology, restricting the flow of information. A sample graph is provided in Fig.~\ref{fig:graph}.
\begin{figure}[!t]
	\centering
	\includegraphics[width=.9\columnwidth]{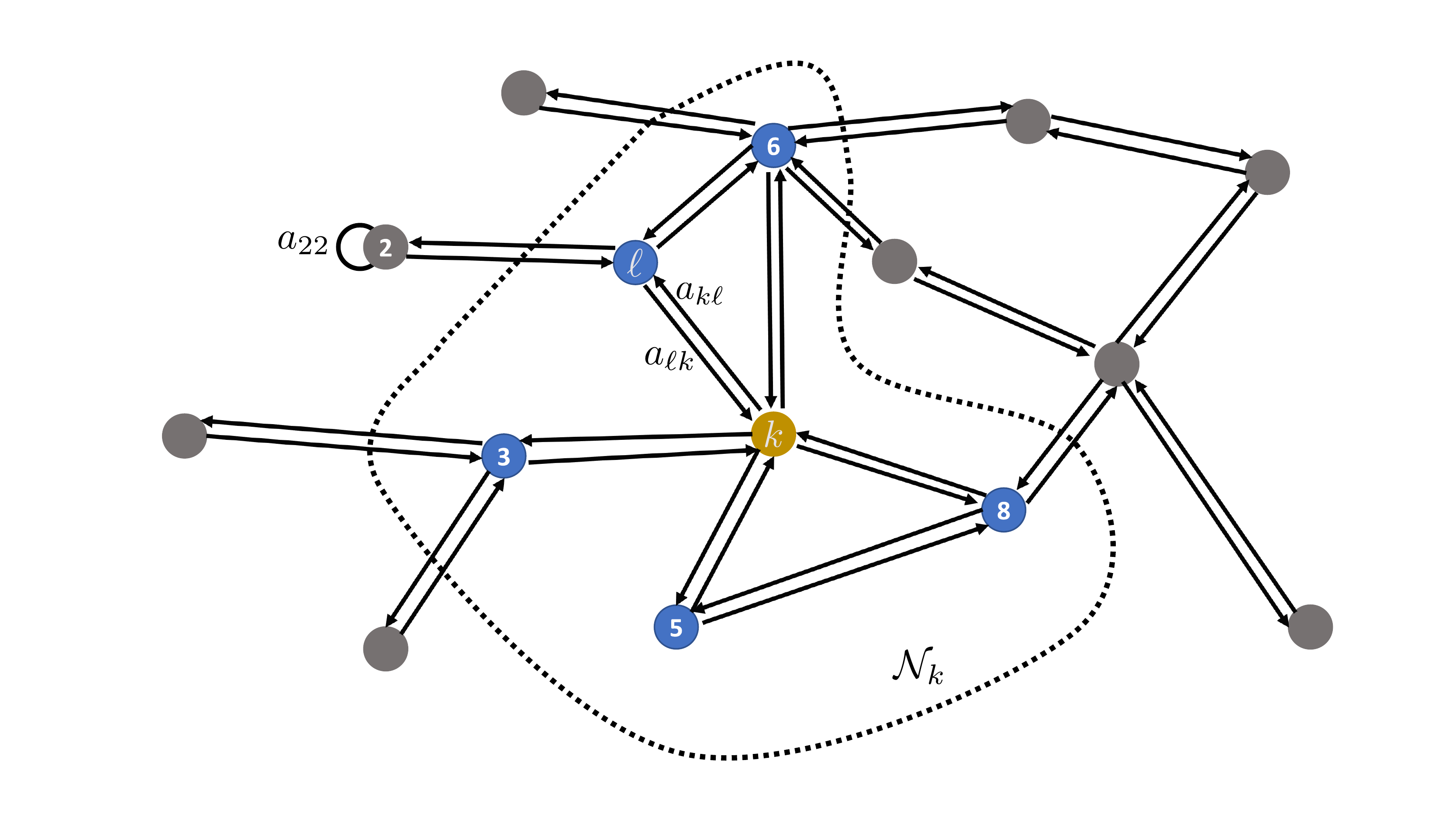}
	\caption{A sample network with an emphasis on the neighborhood \( \mathcal{N}_k \) of node \( k \). Node \( k \) can aggregate information from only its neighbors in \( \mathcal{N}_k \), with \( a_{\ell k} \) denoting the weight given by node \( k \) to information from \( \ell \). {Double-arrows indicate the asymmetric flow of information, since we allow for \( a_{\ell k} \neq a_{k \ell}\).}}\label{fig:graph}
\end{figure}
It is then natural to ask whether the collection of agents can still pursue a solution of~\eqref{eq:decentralized_problem} despite being restricted to performing only local computations and \emph{exchanges of information over neighborhoods}. The answer is indeed affirmative, so long as the network linking the agents is connected, allowing for information to diffuse through the entire network through repeated local aggregations. The solution can then be pursued through a plethora of decentralized strategies, including primal~\cite{Nedic09,  Nedic10, Sayed14, Chen13, Sayed14proc, Chen15transient, DiLorenzo16, Xin18, Yuan19} and (primal-)dual~\cite{Duchi12, Tsianos12, Shi14, Jaggi14, Ling15, Jakovetic15, Jakovetic19} frameworks. A detailed discussion of the properties of these strategies, primarily studied in the convex setting, is beyond the scope of this work. We will instead focus on the diffusion strategy~\cite{Chen13, Sayed14, Sayed14proc}, and discuss its second-order guarantees in nonconvex environments. The diffusion strategy takes the form:
\begin{subequations}
\begin{align}
  \boldsymbol{\phi}_{k,i} &= \w_{k,i-1} - \mu \widehat{\nabla J}_{k}(\w_{k,i-1})\label{eq:adapt}\\
  \w_{k,i} &= \sum_{\ell=1}^{N} a_{\ell k} \boldsymbol{\phi}_{\ell,i}\label{eq:combine}
\end{align}
\end{subequations}
Note that the strategy has an ``adapt-then-combine'' form where in step~\eqref{eq:adapt}, agent \( k \) takes a local descent step using the approximation \( \widehat{\nabla J}_{k}(\w_{k,i-1}) \) based on its locally available data to generate an intermediate estimate \( \boldsymbol{\phi}_{k,i} \). The adaptation step is followed by a combination step where agent \( k \) performs a convex combination of the intermediate estimates \( \boldsymbol{\phi}_{\ell,i} \) using the weights \( a_{\ell k} \) to form \( \w_{k,i} \). We shall make the following assumption on the combination weights.
\begin{assumption}[\textbf{Strongly-connected graph}]\label{as:strongly_connected}
	The graph described by the weighted combination matrix \(A=[a_{\ell k}]\) is strongly-connected~\cite{Sayed14}. This means that there exists a path with nonzero weights between any two agents in the network and, moreover, at least one agent has a nontrivial self-loop, \(a_{kk}>0\). The combination weights satisfy:
  \begin{equation}\label{eq:combinationcoef}
    a_{\ell k} \geq 0, \quad \sum_{\ell \in \mathcal{N}_k} a_{\ell k}=1, \quad a_{\ell k} = 0\ \mathrm{if}\ \ell \notin \mathcal{N}_k
  \end{equation}
  where the symbol \( \mathcal{N}_k \) denotes the set of neighbors of agent \( k \).\hfill\qed%
\end{assumption}
Relation~\eqref{eq:combinationcoef} ensures that~\eqref{eq:combine} indeed carries the interpretation of a convex combination, and can be evaluated by collecting intermediate estimates \( \boldsymbol{\phi}_{\ell,i} \) only from nodes in the immediate neighborhood \( \ell \in \mathcal{N}_{k} \). Strong connectivity of the graph on the other hand, in light of the Perron-Frobenius theorem~\cite{Horn03, Pillai05, Sayed14}, ensures that the combination matrix \( A \) has a single eigenvalue at one with all other eigenvalues strictly within the unit circle. The right eigenvalue of \( A \), denoted by \( p \) can be normalized so that its elements are strictly positive and add up to one~\cite{Sayed14}:
\begin{equation}\label{eq:perron}
  Ap=p, \quad \mathds{1}^{\T} p=1, \quad p_k > 0
\end{equation}
where the \( \{ p_k \} \) denote the individual entries of the Perron vector, \(p\). In the strongly-convex case, it is well known that the diffusion strategy~\eqref{eq:adapt}--\eqref{eq:combine} converges to the minimizer of~\eqref{eq:decentralized_problem} in the mean-square sense where the weights \( p_k \) in~\eqref{eq:decentralized_problem} correspond to the entries of the Perron vector \( p \) in~\eqref{eq:perron} of the combination matrix \( A \)~\cite{Chen13, Sayed14}. It is common to choose \( A \) to be symmetric, resulting in \( p_k = \frac{1}{K} \) for all \( k \) and equal contribution of all nodes in~\eqref{eq:decentralized_problem}. Allowing for more general left- instead of only doubly-stochastic matrices provides the designer with the additional flexibility to, for example, assign larger weight to nodes with higher quality of data, a fact that has been exploited both in the strongly-convex~\cite{Sayed14} and nonconvex~\cite{Vlaski19linearspeedup} setting. In both cases these strategies exploit that for any connected graph, a combination matrix satisfying Assumption~\ref{as:strongly_connected} can be designed in a decentralized manner to have an arbitrary \( p \) as its Perron vector \( Ap=p \)~\cite[Eq. (8.96)]{Sayed14}.

\section{Network Dynamics}
The fact that \( p \) corresponds to a right eigenvector of the combination matrix \( A \) implies for the weighted centroid \( \w_{c, i} \triangleq \sum_{k=1}^K p_k \w_{k, i} \)~\cite{Chen15transient, Sayed14}:
\begin{equation}\label{eq:weighted_recursion}
  \w_{c, i} = \w_{c, i-1} - \mu \sum_{k=1}^K p_k \widehat{\nabla J}_k(\w_{k, i-1})
\end{equation}
Examination of~\eqref{eq:weighted_recursion} shows that \( \w_{c, i} \) evolves \emph{almost} according to a (stochastic) gradient recursion relative to the aggregate cost~\eqref{eq:decentralized_problem} with the subtle difference that the gradient approximations \( \widehat{\nabla J}_k(\w_{k, i-1}) \) are evaluated at the local iterates \( \w_{k, i-1} \) instead of the centroid \( \w_{c, i-1} \). Nevertheless, as long as the collection of iterates \( \w_{k, i-1} \) do not deviate too much from each other, and hence from the (weighted) average \( \w_{c, i} = \sum_{k=1}^K p_k \w_{k, i} \), one would expect the evolution of \( \eqref{eq:weighted_recursion} \) to carry similar performance guarantees to the single-agent and federated solutions~\eqref{eq:sgd_general} and~\eqref{eq:federated_together}, respectively. This has been rigorously established in the strongly-convex case~\cite{Sayed14, Chen15transient, Chen15performance}. In this work, we present more recent extensions to nonconvex risks and second-order guarantees.
\begin{theorem}[\textbf{Network disagreement (4th order)~\cite{Vlaski19nonconvexP1}}]\label{LEM:NETWORK_DISAGREEMENT_FOURTH}
	Suppose each local \( J_k(\cdot) \) and stochastic gradient approximation \( \widehat{\nabla J}_k(\cdot) \) satisfy Assumptions~\ref{as:lipschitz}--\ref{as:bounded} with \( \beta = 0 \). Furthermore, assume the combination matrix \( A \) satisfies Assumption~\ref{as:strongly_connected} with Jordan decomposition \( A = V_{\epsilon} J V_{\epsilon}^{-1} \):
  \begin{equation}\label{eq:jordan}
  V_{\epsilon} = \left[ \begin{array}{cc} p & V_R \end{array} \right],
  \ \ J = \left[ \begin{array}{cc} 1 & 0\\0 & J_{\epsilon} \end{array}\right],
  \ \ V_{\epsilon}^{-1} = \left[ \begin{array}{c} \mathds{1}^{\T} \\ \vphantom{O^{O^{O^O}}} V_L^{\T} \end{array}\right]
\end{equation}
Collect the iterates \( \w_{k, i} \) across the network into \( \bcw_i \triangleq \mathrm{col}\left\{ \w_{1, i}, \ldots, \w_{K, i} \right\} \). Then, the network disagreement is bounded after sufficient iterations \( i \ge i_o \) by:
	\begin{align}
		\E {\left \| \bcw_i - \left( \mathds{1} p^{\T} \otimes I \right) \bcw_{i} \right \|}^4 \le \mu^4 {\left \| \mathcal{V}_L \right \|}^4 \frac{{\left \|J_{\epsilon}^{\T} \right \|}^4}{{\left(1-{\left \|J_{\epsilon}^{\T} \right \|}\right)}^4} {\| \mathcal{V}_R^{\T} \|}^4 N^2 \left( G^4 + \sigma^4 \right) + o(\mu^4)\label{eq:network_disagreement_fourth}
	\end{align}
  where \( \mathcal{V}_L \otimes I_M \), \( \mathcal{V}_R = V_R \otimes I_M \) and
  \begin{equation}
    i_o = \frac{\log\left( o(\mu^4) \right)}{\log\left( {\left \|J_{\epsilon}^{\T} \right \|} \right)}
  \end{equation}
  and \( o(\mu^4) \) denotes a term that is higher in order than \( \mu^4 \).\hfill\qed%
\end{theorem}
To develop some intuition about the implications of~\eqref{eq:network_disagreement_fourth}, observe that we can bound:
\begin{align}
  \frac{1}{N} \sum_{k=1}^K \E \| \w_{k, i} - \w_{c, i}\|^2 =&\: \frac{1}{N}\E {\left \| \bcw_i - \left( \mathds{1} p^{\T} \otimes I \right) \bcw_{i} \right \|}^2 \notag \\
  \stackrel{(a)}{\le}&\: \frac{1}{N}\sqrt{\E {\left \| \bcw_i - \left( \mathds{1} p^{\T} \otimes I \right) \bcw_{i} \right \|}^4} \notag \\
  \stackrel{(b)}{\le}&\: \mu^2 {\left \| \mathcal{V}_L \right \|}^2 \frac{{\left \|J_{\epsilon}^{\T} \right \|}^2}{{\left(1-{\left \|J_{\epsilon}^{\T} \right \|}\right)}^2} {\| \mathcal{V}_R^{\T} \|}^2 N \left( G^2 + \sigma^2 \right) + o(\mu^2)\label{eq:network_disagreement_second}
\end{align}
where \( (a) \) follows from Jensen's inequality along with convexity of \( \|\cdot\|^2 \) and \( (b) \) follows from sub-additivity of the square root \( \sqrt{a+b} \le \sqrt{a} + \sqrt{b} \). We hence conclude that~\eqref{eq:network_disagreement_second} bounds the average deviation of the local iterates \( \w_{k, i} \) from the centroid \( \w_{c, i} \) in the mean-square sense by a term that is on the order of \( \mu^2 \), which is small enough to be negligible for sufficiently small step-sizes \( \mu \). This allows us to derive essentially the same performance guarantees for the network centroid \( \w_{c, i} \) as for the centralized recursion~\eqref{eq:sgd_general}, after accounting for the small and controllable deviation~\eqref{eq:network_disagreement_second}. We make a minor adjustment to the space decomposition from Definition~\ref{DEF:SETS}.
\begin{definition}[Sets]\label{DEF:SETS_DEC}
  We continue with the decomposition into \( \mathcal{G} \), \(\mathcal{H}\) and \(\mathcal{M}\) from relations~\eqref{eq:define_g}--\eqref{eq:define_m} in Definition~\ref{DEF:SETS} and only adjust expression~\eqref{eq:define_c1} for \( c_1 \) to:
  \begin{align}
		c_1 &\triangleq \frac{1}{2} \left( 1 - 2 \mu \delta \right) = O(1) \label{eq:define_c1_dec}
	\end{align}
  and \( 0 < \pi < 1 \) is a parameter to be chosen. Note that \( \mathcal{G}^C = \mathcal{H} \cup \mathcal{M} \). We also define the probabilities \( \pi^{\mathcal{G}}_i \triangleq \mathrm{Pr}\left \{ \w_{c, i} \in \mathcal{G} \right \} \), \(\pi^{\mathcal{H}}_i \triangleq \mathrm{Pr}\left \{ \w_{c, i} \in \mathcal{H} \right \} \) and \(\pi^{\mathcal{M}}_i \triangleq \mathrm{Pr}\left \{ \w_{c, i} \in \mathcal{M} \right \}\). Then for all \( i \), we have \( \pi^{\mathcal{G}}_i + \pi^{\mathcal{H}}_i + \pi^{\mathcal{M}}_i = 1 \).\hfill\qed%
\end{definition}
Note that the only difference between Definitions~\ref{DEF:SETS} and~\ref{DEF:SETS_DEC} is in the definition of \( c_1 \) in~\eqref{eq:define_c1} and~\eqref{eq:define_c1_dec}. This variation is motivated by the technical details of the arguments leading to the descent relations that follow, but ultimately does not change the implications of the result. We then obtain the decentralized versions to the centralized descent Theorems~\ref{LEM:DESCENT_RELATION} through~\ref{TH:FINAL_THEOREM}, established in~\cite{Vlaski19nonconvexP1, Vlaski19nonconvexP2}.
\begin{theorem}[\textbf{Descent relation~\cite{Vlaski19nonconvexP1}}]\label{TH:DESCENT_RELATION_DEC}
	Beginning at \( \w_{c, i-1} \) in the large gradient regime \( \mathcal{G} \), we can bound:
  \begin{align}\label{eq:descent_in_g_dec}
    &\:\E \left \{ J(\w_{c, i}) | \w_{c, i-1} \in \mathcal{G} \right \} \notag \\
    \le&\: \E \left \{ J(\w_{c, i-1}) | \w_{c, i-1} \in \mathcal{G} \right \} - \mu^2 \frac{c_2}{\pi} + \frac{O(\mu^3)}{\pi_{i-1}^{\mathcal{G}}}
  \end{align}
	{as long as \( \pi_{i-1}^{\mathcal{G}} = \mathrm{Pr}\left \{ \w_{c, i-1} \in \mathcal{G} \right \} \neq 0 \)} where the relevant constants are listed in Definition~\ref{DEF:SETS_DEC}. On the other hand, beginning at \( \w_{c, i-1} \in \mathcal{M} \), we can bound:
  \begin{align}\label{eq:ascent_bound_dec}
    &\:\E \left \{ J(\w_{c, i}) | \w_{c, i-1} \in \mathcal{M} \right \} \notag \\
    \le&\: \E \left \{ J(\w_{c, i-1}) | \w_{c, i-1} \in \mathcal{M} \right \} + \mu^2 {c_2} + \frac{O(\mu^3)}{\pi_{i-1}^{\mathcal{M}}}
  \end{align}
  {as long as \( \pi_{i-1}^{\mathcal{M}} = \mathrm{Pr}\left \{ \w_{c, i-1} \in \mathcal{M} \right \} \neq 0 \)}.\hfill\qed%
\end{theorem}
\begin{theorem}[\textbf{Descent through strict saddle-points~\cite{Vlaski19nonconvexP2}}]\label{TH:DESCENT_THROUGH_SADDLE_POINTS_DEC}
  {Suppose \( \mathrm{Pr} \left \{ \w_{c, i} \in \mathcal{H} \right \} \neq 0 \), i.e., \( \w_{c, i} \)} is approximately stationary with significant negative eigenvalue. Then, iterating for \( i^s \) iterations after \( i \) with
  \begin{align}
    i^{s} = \frac{\log\left( 2 M  \frac{\sigma^2}{\sigma_{\ell}^2} + 1 \right)}{\log({1 + 2\mu\tau})} \le O\left(\frac{1}{\mu \tau} \right)
  \end{align}
  guarantees
  \begin{align}
    &\: \E \left \{ J(\w_{c, i+i^s}) | \w_{c, i} \in \mathcal{H} \right \} \notag \\
    \le&\: \E \left \{ J(\w_{c, i}) | \w_{c, i} \in \mathcal{H} \right \} - \frac{\mu}{2} M \sigma^2 + o(\mu) + \frac{o(\mu)}{\pi_{i}^{\mathcal{H}}}
  \end{align}\hfill\qed%
\end{theorem}
\begin{theorem}[\textbf{Second-order guarantee for diffusion~\cite{Vlaski19nonconvexP2}}]\label{TH:FINAL_THEOREM_DEC}
  For sufficiently small step-sizes \( \mu \), we have with probability \( 1 - \pi \), that \( \w_{c, i^o} \in \mathcal{M} \), i.e., \( \| \nabla J(\w_{c, i^o}) \|^2 \le O(\mu) \) and \( \lambda_{\min}\left( \nabla^2 J(\w_{c, i^o}) \right) \ge -\tau \) in at most \( i^o \) iterations, where
  \begin{align}
    i^o \le \frac{\left( J(w_{c, 0}) - J^o \right)}{\mu^2 c_2 \pi} i^s
  \end{align}
  and \( i^s \) denotes the escape time from Theorem~\ref{TH:DESCENT_THROUGH_SADDLE_POINTS_DEC}, i.e.,
  \begin{align}
    i^{s} = \frac{\log\left( 2 M  \frac{\sigma^2}{\sigma_{\ell}^2} + 1 \right)}{\log({1 + 2\mu\tau})} \le O\left(\frac{1}{\mu \tau} \right)
  \end{align}\hfill\qed%
\end{theorem}
Comparing Theorems~\ref{TH:DESCENT_RELATION_DEC}--\ref{TH:FINAL_THEOREM_DEC} to Theorems~\ref{LEM:DESCENT_RELATION}--\ref{TH:FINAL_THEOREM} we note that the descent and second-order stationarity guarantees for the network centroid \( \w_{c, i} \) generated by the diffusion algorithm~\eqref{eq:adapt}--\eqref{eq:combine} are essentially the same as those for the ordinary stochastic gradient descent recursion~\eqref{eq:sgd_general} after adjusting the constants to account for the decentralized nature. Theorem~\ref{LEM:NETWORK_DISAGREEMENT_FOURTH}, on the other hand, ensures that all iterates \( \w_{k, i} \) will closely track the network centroid \( \w_{c, i} \) after sufficient iterations, and hence each agent \( k \) in the network with inherit the second-order guarantees of \( \w_{c, i} \).

\section{Simulation Example}
We illustrate the theoretical results in this work on a simple example, motivated by neural network learning and used as a benchmark in~\cite{Vlaski19nonconvexP1, Vlaski19nonconvexP2, Vlaski19single}. Given a feature vector \( \boldsymbol{h} \in \mathds{R}^M \) and binary label \(\boldsymbol{\gamma} \in \left\{ 0, 1\right\}\), we model a learning rule \( \widehat{\gamma}\left( \boldsymbol{h}; \cdot \right) \) through a neural network with one linear hidden layer and a logistic activation function at the output layer, taking the form:
\begin{equation}
  \widehat{\gamma}\left( \boldsymbol{h}; w_1, W_2 \right) \triangleq \frac{1}{1+e^{-w_1^{\T} W_2 \boldsymbol{h}}}
\end{equation}
where \( w_1 \in \mathds{R}^N, W_2 \in \mathds{R}^{N\times M} \) denote the model parameters. We employ the cross-entropy loss:
\begin{equation}
  Q(w_1, W_2; \boldsymbol{h}, \boldsymbol{\gamma}) = - \boldsymbol{\gamma}\log\left(\widehat{\gamma}(\boldsymbol{h}; w_1, W_2)\right) - (1-\boldsymbol{\gamma})\log\left(1-\widehat{\gamma}(\boldsymbol{h}; w_1, W_2)\right)
\end{equation}
The risk obtained after taking expectations and adding regularization is:
\begin{equation}
  J(w_1, W_2) \triangleq \E Q(w_1, W_2; \boldsymbol{h}, \boldsymbol{\gamma}) + \frac{\rho}{2} \|w_1\|^2 + \frac{\rho}{2} \|W_2\|^2
\end{equation}
This risk function is nonconvex, and for \( M = N = 1 \) has two local minima in the positive and negative quadrants, respectively with a single strict saddle-point at \( w_1 = W_2 = 0 \). This risk surface was used in Figures~\ref{fig:space_decomp} and~\ref{fig:alignment} to illustrate the modeling conditions for the theorems in this work. We now illustrate the practical performance of the centralized strategy~\eqref{eq:sgd_general}, the federated algorithm~\eqref{eq:federated_together}, as well as the decentralized diffusion strategy~\eqref{eq:adapt}--\eqref{eq:combine} and verify that the second-order guarantees established in Theorems~\ref{TH:FINAL_THEOREM} and~\ref{TH:FINAL_THEOREM_DEC} indeed hold.

We consider a collection of \( K = 50 \) agents, each sampling independent pairs \( \left\{ \boldsymbol{h}_k, \boldsymbol{\gamma}(k) \right\} \) once per iteration. Although it is not a requirement of the analysis, for simplicity, we consider in this example a homogeneous data setting, where pairs \( \left\{ \boldsymbol{h}_k, \boldsymbol{\gamma}(k) \right\} \) are identically distributed for all agents and \( J_k(w_1, W_2) = J(w_1, W_2) \) for all \( k \). The agents are linked by a random graph with mixing rate \( \rho\left( A - \mathds{1}p^{\T} \right) = 0.956 \) with combination weights giving equal weight to all neighbors. We collect \( w \triangleq \mathrm{col}\left\{ w_1, \mathrm{vec}\left\{ W_2 \right\} \right\} \). For the centralized solution~\eqref{eq:sgd_general}, the stochastic gradient approximation is constructed by aggregating data from \emph{all} agents:
\begin{equation}
  \widehat{\nabla J}(w) \triangleq \sum_{k=1}^{K} p_k \nabla Q(w; \boldsymbol{h}_k, \boldsymbol{\gamma}(k)) + \rho w + \boldsymbol{v} \cdot \mathrm{col}\left\{ 1, 1 \right\}
\end{equation}
where the random perturbation \( \boldsymbol{v} \sim \mathcal{N}(0, 1) \) was added to ensure that Assumption~\ref{as:noise_in_saddle} holds. For the federated implementation~\eqref{eq:federated_together}, each of the \( L = 10 \) participating agents at each iteration construct:
\begin{equation}
  \w_{k, i} = \w_i - \mu K p_k \mathds{1}_{k, i} \left( \left( \nabla Q(w; \boldsymbol{h}_k, \boldsymbol{\gamma}(k)) + \rho w + \boldsymbol{v} \cdot \mathrm{col}\left\{ 1, 1 \right\} \right) \right)
\end{equation}
The intermediate estimates \( \w_{k, i} \) are then fused according to~\eqref{eq:federated_combination}. For the decentralized implementation, each agent constructs:
\begin{equation}
  \widehat{\nabla J}_k(w) \triangleq \nabla Q(w; \boldsymbol{h}_k, \boldsymbol{\gamma}(k)) + \rho w + \boldsymbol{v} \cdot \mathrm{col}\left\{ 1, 1 \right\}
\end{equation}
and then updates according to~\eqref{eq:adapt}--\eqref{eq:combine}. All iterates for all three strategies are initialized at \( \left\{ 0.8, -0.8 \right\} \). As predicted by the theoretical results, all three strategies are able to escape the saddle-point at \( w_1 = W_2 = 0 \). Detailed performance is shown in Fig.~\ref{fig:simulation}.
\begin{figure}
\centering
\begin{subfigure}{.5\textwidth}
  \centering
  \includegraphics[width=\linewidth]{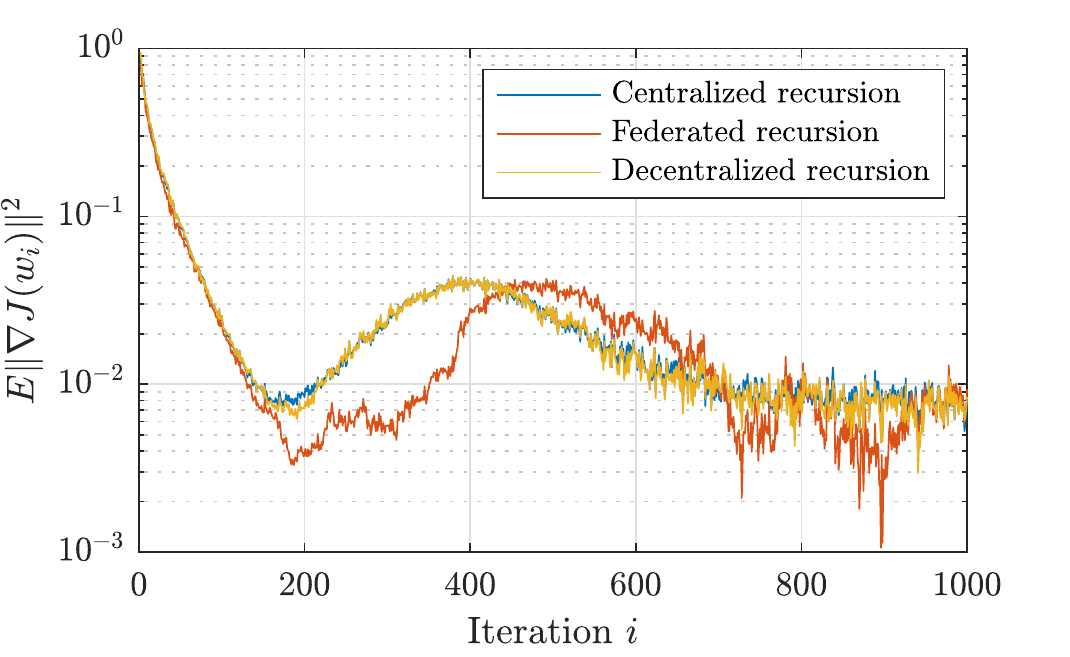}
\end{subfigure}%
\begin{subfigure}{.5\textwidth}
  \centering
  \includegraphics[width=\linewidth]{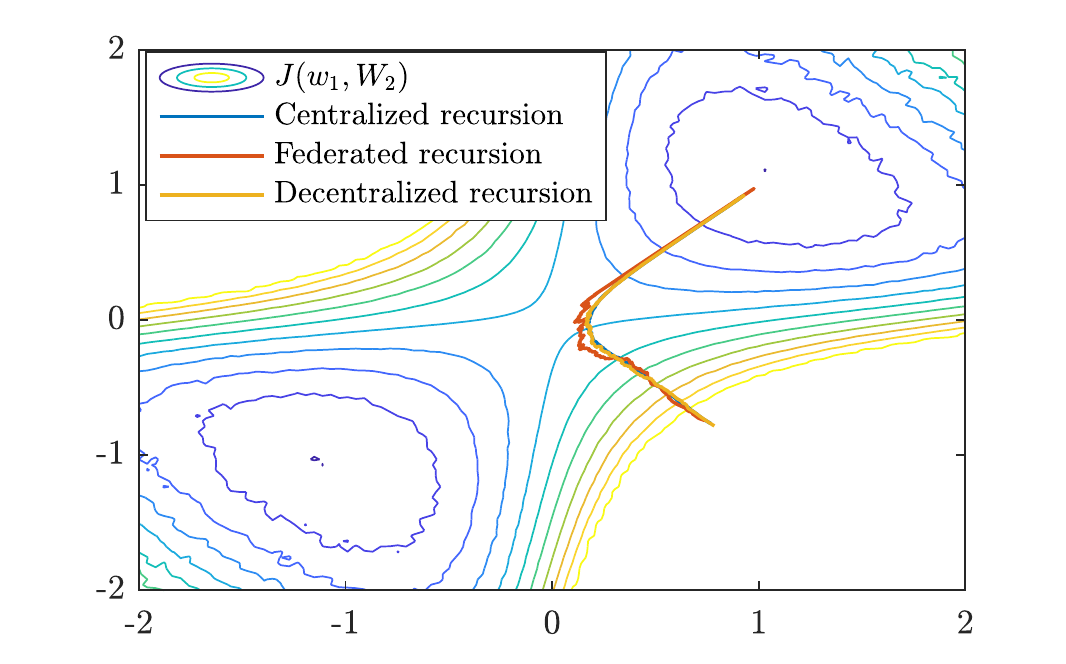}
\end{subfigure}
\caption{(left) Evolution of the gradient norm. (right) Evolution of the iterates. Around iteration \( 200 \), all three algorithms approach the saddle-point, where the norm of the gradient drops and the evolution slows. Due to the presence of gradient perturbations, all three algorithms are able to escape from the saddle-point and eventually reach a local minimum. The decentralized solution~\eqref{eq:adapt}--\eqref{eq:combine} closely tracks the centralized algorithm~\eqref{eq:sgd_general}, while the federated algorithm~\eqref{eq:federated_together} exhibits slightly higher variance due to the scaled step-size to account for partial agent participation.}
\label{fig:simulation}
\end{figure}

\section{Conclusion}
In this manuscript we presented recent results from~\cite{Vlaski19single, Vlaski19nonconvexP1, Vlaski19nonconvexP2} establishing second-order guarantees for stochastic descent algorithms in centralized, federated, and decentralized settings. Two key conclusions emerge. First, we found that in all cases, simple first-order descent algorithms are able to yield second-order optimal solutions, which exclude saddle-points and correspond to local or even global minima in many problems of interest, so long as their recursions are subjected to sufficient perturbations. These perturbations are critical in ensuring that the recursions do not spend extraordinary amounts of time near saddle-points, where the progress of unperturbed gradient recursions is slow~\cite{Du17}. Second, under a reasonable bound on agent heterogeneity, we found that for sufficiently small step-sizes, the performance guarantees of the decentralized strategy essentially match those for the centralized framework, implying that even in the absence of central aggregation of data or parameter estimates, decentralized strategies can yield competitive performance in terms of their second-order guarantees, a fact that is well established for strongly-convex costs, but only recently has received attention in the nonconvex setting.

\bibliographystyle{IEEEbib}
\bibliography{main}

\end{document}

%% file: custom/operators.tex
\usepackage{amsthm}

\DeclareMathOperator{\bcw}{{\boldsymbol{\scriptstyle\mathcal{W}}}}
\DeclareMathOperator*{\argmin}{arg\,min}

\DeclareMathOperator{\T}{\mathsf{T}}
\DeclareMathOperator{\E}{\mathds{E}}

\DeclareMathOperator{\w}{\boldsymbol{w}}
\DeclareMathOperator{\x}{\boldsymbol{x}}
\DeclareMathOperator{\s}{\boldsymbol{s}}

\DeclareMathOperator{\h}{\boldsymbol{h}}

%% file: custom/environments.tex
\usepackage{amsthm}

\theoremstyle{definition}
\newtheorem{example}{Example}
\theoremstyle{plain}
\newtheorem{definition}{Definition}
\newtheorem{assumption}{Assumption}
\newtheorem{theorem}{Theorem}
\newtheorem{corollary}{Corollary}